\documentclass[aps,pre,reprint,unsortedaddress]{revtex4-1}
\usepackage[version=3]{mhchem}
\usepackage{amsmath}
\usepackage{pgfplots}
\usepackage{float}
\usepackage[hidelinks]{hyperref}
\usepackage{enumerate}
\usepackage{footnote}
\usepackage{perpage} 
\MakePerPage{footnote} 
\usepackage{booktabs}
\usepackage{changepage} 
\usepackage{siunitx}
\usepackage{systeme,mathtools}
\usepackage{verbatim}
\usepackage{graphicx, color}

\usepackage[utf8]{inputenc}
\usepackage[T1]{fontenc}
\usepackage{breqn}
\usepackage{textcomp}
\usepackage{gensymb}

\begin{document}
\title{The Heat of Nervous Conduction$:$ A Thermodynamic Framework}

\author{Aymar C. L. de Lichtervelde$^{1,\dagger}$}
\author{J. Pedro de Souza$^{2,\dagger}$}%
\author{Martin Z. Bazant$^{2,3, *}$}

\affiliation{$^{1}Department\:of\:Physical\:Chemistry\:\&\:Soft Matter,\break Wageningen\:University,\:6708\:WG\:Wageningen, \:The\:Netherlands$}
\affiliation{$^2\:Department\:of\:Chemical\:Engineering\:and\:^{3} Department\:of\:Mathematics,\break Massachusetts\:Institute\:of\:Technology, Cambridge,\:MA\:02139,\:USA$}%

\date{\today}
\begin{abstract}
Early recordings of nervous conduction revealed a notable thermal signature associated with the electrical signal. The observed production and subsequent absorption of heat arise from physicochemical processes that occur at the cell membrane level during the conduction of the action potential. In particular, the reversible release of electrical energy stored as a difference of potential across the cell membrane appears as a simple yet consistent explanation for the heat production, as proposed in the ``Condenser Theory.'' However, the Condenser Theory has not been analyzed beyond the analogy between the cell membrane and a parallel-plate capacitor, \textit{i.e.} a condenser, which cannot account for the magnitude of the heat signature.
In this work, we use a detailed electrostatic model of the cell membrane to revisit the Condenser Theory. We derive expressions for free energy and entropy changes associated with the depolarization of the membrane by the action potential, which give a direct measure of the heat produced and absorbed by neurons. We show how the density of surface charges on both sides of the membrane impacts the energy changes. 
Finally, considering a typical action potential, we show that if the membrane holds a bias of surface charges, such that the internal side of the membrane is 0.05 C m$^{-2}$ more negative than the external side, the size of the heat predicted by the model reaches the range of experimental values. Based on our study, we identify the change in electrical energy of the membrane as the primary mechanism of heat production and absorption by neurons during nervous conduction. 
\end{abstract}
\maketitle

\section{Introduction}
\subsection{Thermodynamics of nervous conduction}
Besides the electrical responses classically measured in electrophysiological experiments \cite{hodgkin1952measurement,hodgkin1952components,neher1992,hamill1981}, the action potential is accompanied with a production and a subsequent absorption of heat \cite{Howarth1979}, changes in optical properties \cite{Tasaki1968}, and mechanical deformations \cite{Tasaki1989,Gonzalez-Perez2016}.  
These thermal, optical, and mechanical responses are macroscopic signatures of the physicochemical processes occurring at the cell membrane level during the action potential, such as the transport of ions through ion channels in the membrane \cite{Hille2001} or the elastic deformation of the membrane \cite{ElHady2015}. While such physicochemical signatures are associated with the electrical signal, classical electrical circuit models cannot accurately capture them because the circuit models neglect the microscopic physics at the membrane level.

Here we examine the thermal response of nervous conduction, by resolving the microscopic physics of the membrane and its surrounding electrical double layers. 
We start by reviewing experimental and theoretical backgrounds on the heat signature of nervous conduction. Then we apply an equilibrium-thermodynamics framework to calculate the electrical energy that is stored by the membrane and the surrounding double layers and released into heat during the passage of the action potential. Finally, based on typical neurophysiological parameters, we show that the reversible release of electrical energy offers a plausible explanation for the heat of nervous conduction.

\subsection{Heat production and absorption by neurons: the experimental context}
A substantial experimental record shows that the propagation of the action potential along neurons is accompanied by the release of a small amount of heat, immediately followed by the absorption of a similar amount of heat by the neurons \cite{Abbott1958,Howarth1968,Ritchie1985,Howarth1979}.
Successfully measured for the first time in 1925 \cite{Ritchie1973}, the heat of nervous conduction has been most extensively investigated between the 1950s and 1980s by contemporaries and colleagues of Hodgkin and Huxley \cite{Abbott1958,Howarth1968,Ritchie1985,Howarth1979}, the pioneers of modern neurophysiology. All neurons possess a similar excitable membrane, and heat production is likely a universal feature of nervous conduction. However, the thermal signals are most easily measured in nerve fibers that have a high surface-to-volume ratio. 
The thermal signals are indeed extremely small, and it appears that the heat flux is proportional to the axon membrane area \cite{Howarth1979}. The garfish olfactory nerve, for example, is an excellent candidate for recording the heat of nervous conduction: it is made of several millions of fibers of 0.25 $\SI{}{\micro\meter}$ in diameter, totals a membrane area of 6.5 m$^2$ per g of nerve \cite{easton1971}, and releases heat on the order of 1 mJ g$^{-1}$ \cite{Howarth1979}. When expressed per total membrane area, the size of the heat remains on the same order of magnitude from one organism to the other (60 - 180 $\SI{}{\micro\joule}$ m$^{-2}$ \cite{Howarth1979}). 

To understand the origin of the thermal signals, scientists attempted to correlate them with the electrical signals \cite{Abbott1958,Howarth1968,Howarth1979}. Notably, \citet{Howarth1979} successfully reconstructed the true temperature change that occurs in neurons from recorded heat responses, and showed that the time course of the temperature changes closely matches the one of the membrane potential during the action potential. Such finding gave support to the ``Condenser Theory.''

\subsection{The Condenser Theory}
The Condenser Theory offers a simple explanation for the heat production and absorption: it attributes them to the reversible release of electrical energy stored across the cell membrane \cite{Abbott1958,Howarth1968,Howarth1979,Ritchie1985}. At rest, the membrane of neurons holds a difference of electric potential, called ``membrane potential.'' An action potential occurs when the membrane potential at a specific location rapidly rises (depolarization) and falls (repolarization), due to the opening of ion channels \cite{Hille2001}. The Condenser Theory states that as the action potential depolarizes
the membrane, the electrical energy stored across the cell membrane is released into heat. Conversely, upon repolarization of the membrane to its resting potential, the membrane’s electrical energy is restored at the expense of some of the thermal energy of the ions in the surrounding solutions, which accounts for the heat absorption phase, in symmetry with the production phase. The membrane is seen as a capacitor (or a "condenser"), hence this explanation for the heat of nervous conduction is known as the Condenser Theory. 
In the first developments of the Condenser Theory, the amount of heat reversibly exchanged between the membrane and its surroundings was calculated as the free energy of a parallel-plate capacitor \cite{Howarth1979,Ritchie1985}: 
\begin{equation}\label{eq:DF}
\Delta{F} = \frac{1}{2} c_{\mathrm{m}} ({V_{\mathrm{m}}}^2-{V_{\mathrm{m},0}}^2)
\end{equation} where $\Delta{F}$ is the free energy change (J m${^{-2}}$), $c_{\mathrm{m}}$ (F m${^{-2}}$) the membrane capacitance, $V_{\mathrm{m}}$ the membrane potential (the inside potential minus the outside potential, \textit{cf.} Fig. \ref{fig:Membrane_heat3}) and $V_{\mathrm{m},0}$ the resting potential (typically $- 70$ mV \cite{weiss1996cellular}).
\subsection{Arguments to explain the missing heat}
Though in qualitative agreement with the experimental records, Eq. \eqref{eq:DF} only predicts a quarter of the heat that is measured.
Realising this, \citet{Howarth1979} and \citet{Ritchie1985} suggested that the free energy should be calculated based on the local value of potential falling on each side of the membrane (\textit{cf.} $\phi_{\mathrm{t}}$ in Fig. \ref{fig:Membrane_heat3}), rather than on the potential values in the internal and external bulk solutions (\textit{cf.} $V_{\mathrm{m}}$ in Fig. \ref{fig:Membrane_heat3}):
\begin{equation}\label{eq:DF_Ritchie}
\Delta{F}_{Ritchie} = \frac{1}{2} \:c_{\mathrm{m}}  ({\phi_{\mathrm{t}}}^2-{\phi_{\mathrm{t},0}}^2)
\end{equation}
In particular, it was pointed out that the presence of an uneven distribution of negative surface charges on the membrane (more charges on the internal side than on the external side) would increase the transmembrane potential difference and lead to more heat being evolved \cite{Howarth1979,Ritchie1985}. 
Unfortunately, \citet{Howarth1979} and \citet{Ritchie1985} did not provide a careful derivation for Eq. \eqref{eq:DF_Ritchie}, nor did they explore the physics of the cell membrane and its surface charges beyond the analogy with a parallel-plate capacitor.  

A second argument invoked by several authors to bridge the gap between predicted and measured heats concerns entropy changes presumed to occur inside the lipid bilayer (\textit{i.e.} the membrane) when the electric field across the membrane relaxes \cite{Howarth1968,Ritchie1973,Howarth1979,Ritchie1985}. Specifically, they proposed that the total energy change associated with the depolarization of the membrane could differ significantly from the free energy change $\Delta{F}$, by an entropy contribution $T\Delta{S}$. These changes were calculated proportionally to the free energy term, based on the temperature dependence of the membrane capacitance: 
\begin{equation}\label{eq:entropy-free_energy1}
T\Delta{S} = \Delta{F}\:\frac{T}{c_{\mathrm{m}}}\:\frac{\partial{c_{\mathrm{m}}}}{\partial{T}}
\end{equation}
where $\Delta{S}$ is the entropy change and $T$ the temperature. Eq. \eqref{eq:entropy-free_energy1} predicts an additional heating of the lipid membrane if $T\Delta{S}$ and $\Delta{F}$ have the same sign, otherwise it predicts a cooling. According to \citet{Ritchie1985} and Ref. \cite{taylor1962}, the value of ${T/c_{\mathrm{m}}}\:{\partial{c_{\mathrm{m}}}}/{\partial{T}}$ is positive, between 2 and 4, which would bring a total warming that is 3 to 5 times higher than if the heat was derived only from the release of free energy stored in the membrane capacity. However, no rigorous derivation for Eq. \ref{eq:entropy-free_energy1} could be found in literature. Furthermore, the prediction of \citet{Ritchie1985} seems difficult to reconcile with recent measurements of the temperature dependence of the dielectric permittivity (${\partial{\varepsilon}}/{\partial{T}}$) of fatty acids, the carbon chains that form the cell membrane. Indeed, ${\partial{\varepsilon}}/{\partial{T}}$ appears to be negative \cite{Plaksin2018,Lizhi2008}, and if we calculate the membrane capacitance $c_{\mathrm{m}}$ as proportional to its dielectric permittivity $\varepsilon_{\mathrm{m}}$ we expect ${\partial{c_{\mathrm{m}}}}/{\partial{T}}$ to be negative rather than positive, which contradicts the argument of \citet{Ritchie1985}. 

More than a simple parallel-plate capacitor, the axon membrane consists of a lipid bilayer with surface charges and electrical double layers forming on each side \cite{Genet2000}. The analogy with a condenser offers a too limited description of the membrane to verify the correctness of the arguments reviewed above. It does not allow to judge which of Eqs. \eqref{eq:DF} or \eqref{eq:DF_Ritchie} describe correctly the free energy change during the depolarization of the membrane.
In addition, as shown above, how entropy changes inside the membrane could contribute to the heat production still needs to be understood. 

To assess whether electrical energy changes constitute a plausible explanation for the heat production and absorption by neurons, we will now derive the changes in electric free energy and entropy that accompany the action potential, based on a detailed electrostatic model of the membrane, its surface charges, and double layers.
\section{Theory}
\subsection{Electrostatic model of the charged lipid bilayer}
\begin{figure}[H]
\centering
\includegraphics[width=1\linewidth]{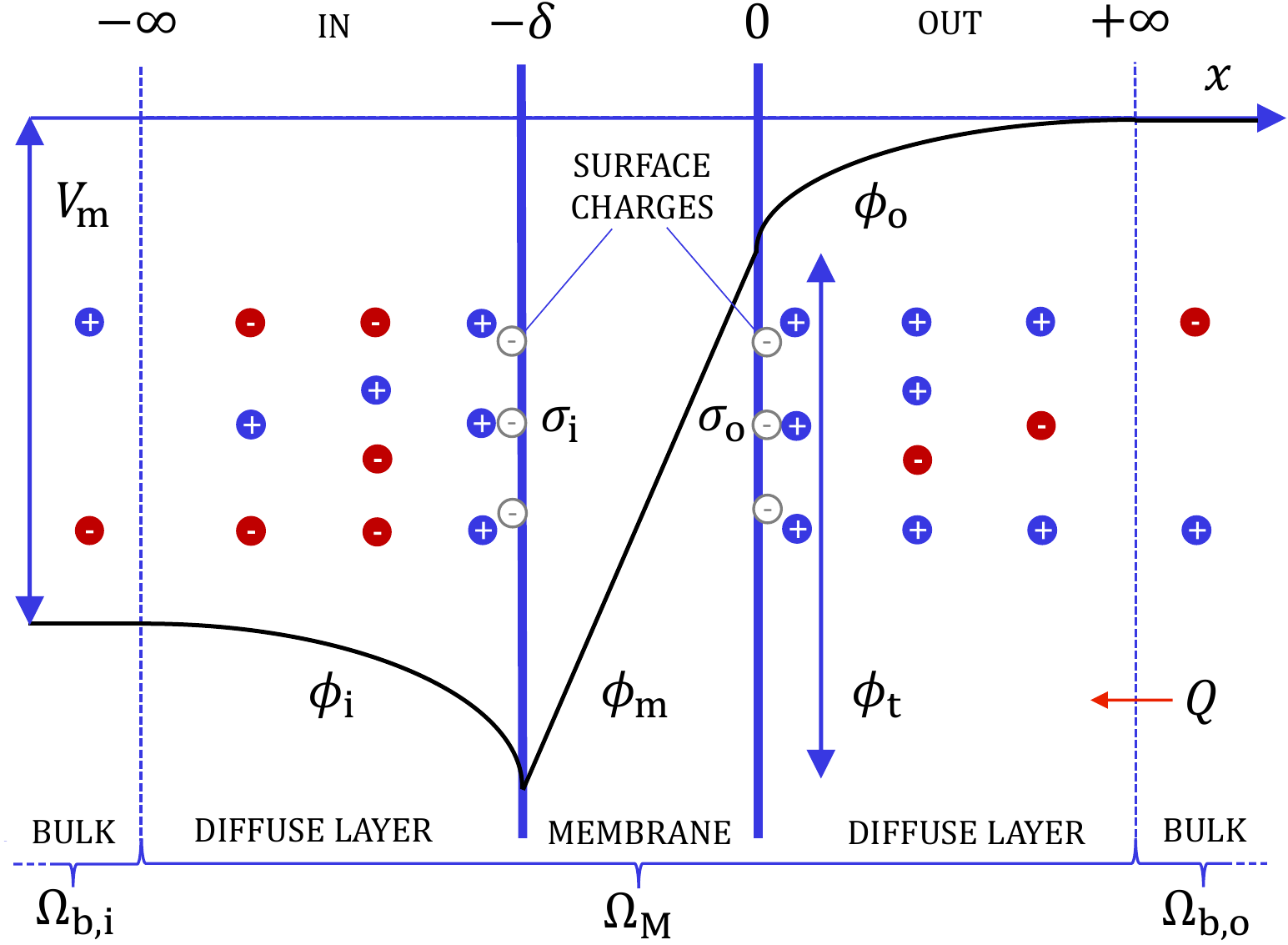}
\caption{Profile of the electric potential across the cell membrane and the diffuse layers on each side. The minus signs in white denote surface charges ($\sigma_{\mathrm{i}}$ inside and $\sigma_{\mathrm{o}}$ outside), while the charges in solution are drawn in blue and red. The dotted lines delimit thermodynamic domains (see Sec. \ref{sec:thermo}). At rest, the membrane holds a potential difference ($V_{\mathrm{m}}=-70$ mV). Upon excitation, cations cross the membrane through ion channels (not depicted here) and $V_{\mathrm{m}}$ jumps to positive values. Simultaneously, electrical energy is dissipated as heat ($Q<0$).
}
\label{fig:Membrane_heat3}
\end{figure}

We use the coupled electrostatic model proposed by \citet{Genet2000}, which applies Poisson-Boltzmann theory on either side of the cell membrane. Fig. \ref{fig:Membrane_heat3} shows the qualitative electric potential profile surrounding and inside a cell membrane that holds surface charges on the internal and external sides ($\sigma_{\mathrm{i}} $ and $\sigma_{\mathrm{o}} $, respectively). 
By convention, we use the symbols $-\infty$ and $+\infty$ to denote the (arbitrary) limits between the diffuse layer and bulk regions, in the internal and external solutions, respectively. Note that in physiological conditions, diffuse layers extend over a few nano-meters at most (the Debye length is 0.6 nm). The membrane core is located between $x=-\delta$ and $x=0$, where $\delta$ is the thickness of the membrane. The $\phi_{\mathrm{i}}$, $\phi_{\mathrm{m}}$ and $\phi_{\mathrm{o}}$ variables represent the potential in the internal solution, membrane core and external solution, respectively. 
The \textit{membrane potential} is defined as the difference between the potential in the internal and external bulk solutions, $V_{\mathrm{m}} = \phi(-\infty) - \phi(+\infty)$, whereas the \textit{transmembrane potential} $\phi_{\mathrm{t}}$ is the potential difference between the internal and external surfaces of the membrane, $\phi_{\mathrm{t}} = \phi_{\mathrm{i}}(-\delta) - \phi_{\mathrm{o}}(0)$. Note that due to the presence of surface charges on each side of the membrane, the local difference of potential $\phi_{\mathrm{t}}$ can differ significantly from the membrane potential $V_{\mathrm{m}}$.
We assume that equilibration of diffuse layers with the bulk electrolytes is fast compared to the dynamics of the action potential (this is verified in Section \ref{sec:fast-eq}), which allows us to describe the concentration of ions in the diffuse layers close to the membrane with the Boltzmann distribution
\begin{align}
    c_{\mathrm{i},j}(x) &= c_{\mathrm{i},j}(b)\:\exp{\{-\frac{z_j\:F\:\phi_{\mathrm{i}}}{R\:T}\}},\\
    c_{\mathrm{o},j}(x) &= c_{\mathrm{o},j}(b)\:\exp{\{-\frac{z_j\:F\:(\phi_{\mathrm{o}}-V_{\mathrm{m}})}{R\:T}\}},
\end{align}
where $c_{\mathrm{i},j}$ and $c_{\mathrm{o},j}$ are the concentrations of species $j$ inside and outside, respectively, and $RT/F$ is the thermal voltage ($RT/F = 23.5$ mV at $T = 273$ K). 
Applying Poisson's equation to each compartment gives
\begin{align}
        \frac{d^2{\phi_{\mathrm{i}}}}{dx^2} &= -\frac{\rho_{\mathrm{i}}}{\varepsilon},\\ 
        \frac{d^2{\phi_{\mathrm{m}}}}{dx^2} &= 0, \\
        \frac{d^2{\phi_{\mathrm{o}}}}{dx^2} &= -\frac{\rho_{\mathrm{o}}}{\varepsilon}, 
\end{align}
where $\rho$ is the density of free charges (C m$^{-3}$), and $\varepsilon$ the dielectric permittivity of the internal and external solutions (water). Note that we assume that the free charge density is zero inside the membrane ($\rho_{\mathrm{m}} = 0$), with a zero ion concentration inside the membrane. 
As $\rho_{\mathrm{i}} = \sum{z_j \: F\:c_{\mathrm{i},j}} $ and $\rho_{\mathrm{o}} = \sum{z_j \:F\:c_{\mathrm{o},j}} $, we obtain the following Poisson-Boltzmann equations:
\begin{align}\label{eq:PBi}
    \frac{d^2\phi_{\mathrm{i}}}{dx^2}(x) &= -\frac{F}{\varepsilon}\sum^N_j\: z_j \:c_{\mathrm{i},j}(b)\:\exp{\{-\frac{z_j F\:\phi_{\mathrm{i}}(x)}{R\:T}\}},\\
    \frac{d^2\phi_{\mathrm{o}}}{dx^2}(x) &= -\frac{F}{\varepsilon}\sum^N_j\:z_j\: c_{\mathrm{o},j}(b)\:\exp{\{-\frac{z_j F\:[\phi_{\mathrm{o}}(x) - V_{\mathrm{m}}]}{R\:T}\}},\label{eq:PBo}
\end{align}
which we solve numerically, using boundary conditions given by Maxwell's equations at an interface (Eqs. \eqref{eq:Maxwell_BCi} and \eqref{eq:Maxwell_BCo}). We now report the electrostatic relations necessary to derive energy changes in following sections. At rest, the membrane acts as a dielectric medium (ion channels are closed), storing a capacitive charge that we define as $-q$ in the internal solution and $+q$ in the external solution. In each compartment, this capacitive charge can be expressed as the sum of the charge that counter-balances the charges that belong to the surface of the membrane (that is $-\sigma_{\mathrm{i}}$ inside and $-\sigma_{\mathrm{o}}$ outside, both with units C m$^{-2}$) and the total mobile charge in solution, which we obtain by integrating the free charge density over the diffuse layers:
\begin{align}\label{eq:rho_i}
- q&=\sigma_{\mathrm{i}}+\int_{-\infty}^{-\delta}{\rho_{\mathrm{i}}} \:dx,   \\
q&=\sigma_{\mathrm{o}} +\int_{0}^{\infty}{\rho_{\mathrm{o}}}\: dx.  
\end{align}
Applying Poisson's equation to the integrals above, we can relate the slope of the membrane potential at each interface to the total charge density in the internal and external solutions
\begin{align}\label{eq:18}
     \frac{d\phi_{\mathrm{i}}}{dx}(-\delta) &= \frac{1}{\varepsilon}(\sigma_{\mathrm{i}} + q), \\
     \frac{d\phi_{\mathrm{o}}}{dx}(0) &= - \frac{1}{\varepsilon}\:\Big(\sigma_{\mathrm{o}} -q\Big),
     \label{eq:19}
\end{align}
Boundary conditions for electric potential at the two membrane-solution interfaces are given by
\begin{align}\label{eq:Maxwell_BCi}
\varepsilon \frac{d\phi_{\mathrm{i}}}{dx}(-\delta) -\varepsilon_{\mathrm{m}}\frac{d\phi_{\mathrm{m}}}{dx}(-\delta)   &= \sigma_{\mathrm{i}},\\
\varepsilon_{\mathrm{m}} \frac{d\phi_{\mathrm{m}}}{dx}(0) -\varepsilon\frac{d\phi_{\mathrm{o}}}{dx}(0)  &= \sigma_{\mathrm{o}},
\label{eq:Maxwell_BCo}
\end{align}
where $\varepsilon_{\mathrm{m}}$ is the dielectric permittivity of the membrane (F m$^{-1}$). By substituting Eq. \eqref{eq:18} into \eqref{eq:Maxwell_BCi}, and Eq. \eqref{eq:19} into \eqref{eq:Maxwell_BCo} we can deduce that
\begin{equation}\label{eq:24}
\frac{q}{\varepsilon_{\mathrm{m}}} = \frac{d\phi_{\mathrm{m}}}{dx}(-\delta) = \frac{d\phi_{\mathrm{m}}}{dx}(0). 
\end{equation}
As the electric field is assumed to be constant inside the membrane, Eq. \eqref{eq:24} gives
\begin{equation}\label{eq:phi_t}
  \frac{q}{\varepsilon_{\mathrm{m}}} = \frac{\phi_{\mathrm{m}}(0) - \phi_{\mathrm{m}}(-\delta)}{\delta} \overset{\Delta}{=} -\frac{\phi_{\mathrm{t}}}{\delta}.
\end{equation}
We finally obtain an expression that relates the capacitive charge to the membrane's capacitance ($c_{\mathrm{m}}$, in F m$^{-2}$) and the transmembrane potential: 
\begin{equation}\label{eq:q}
q = - \frac{\varepsilon_{\mathrm{m}}}{\delta}\: \phi_{\mathrm{t}} = -c_{\mathrm{m}}\:\phi_{\mathrm{t}},
\end{equation}
with $c_{\mathrm{m}} \overset{\Delta}{=} {\varepsilon_{\mathrm{m}}}/{\delta}$, identical to the result from \citet{Genet2000}.
%
\subsection{Thermodynamic definitions}\label{sec:thermo}
To relate electrical energy changes at the membrane level to the exchange of heat with the surroundings, we need to define the membrane as a thermodynamic system. We divide the space into three domains (see Fig. \ref{fig:Membrane_heat3}):  $\Omega_{\mathrm{M}}$ is the ``membrane domain'', \textit{i.e.} the region comprised of the membrane and the diffuse layers that form in the internal and external solutions, while $\Omega_{\mathrm{b,i}}$ and $\Omega_{\mathrm{b,o}}$ denote the internal and external bulk solutions, respectively. The first law of thermodynamics, $\Delta{U_{\mathrm{el}}} = Q + W_{\mathrm{el}}$, applies, in which $\Delta{U_{\mathrm{el}}}$ is the internal energy change associated with the variation of the membrane potential from $V_{\mathrm{m},0}$ to $V_{\mathrm{m}}$, $Q$ is the heat added to $\Omega_{\mathrm{M}}$ ($Q<0$ when heat is dissipated) and $W_{\mathrm{el}}$ is the electrical work done on $\Omega_{\mathrm{M}}$ by the surroundings. Since the capacitive charges $+q$ and $-q$ are confined to the diffuse layers close to the membrane (the bulk solutions are electroneutral), no electrical work is done on $\Omega_{\mathrm{M}}$ by the surrounding domains $\Omega_{\mathrm{b,i}}$ and $\Omega_{\mathrm{b,o}}$. 
The first law thus becomes
\begin{equation}\label{eq:dU_Q}
\Delta{U_{\mathrm{el}}} = Q.
\end{equation}
The internal energy of the membrane system can be expressed as a sum of free energy and entropy-related energy:
\begin{equation}\label{eq:dUdef}
\Delta{U_{\mathrm{el}}} = \Delta{F_{\mathrm{el}}} + T\Delta{S_{\mathrm{el}}},
\end{equation}
where $\Delta{F_{\mathrm{el}}}$ and $\Delta{S_{\mathrm{el}}}$ are respectively the free energy and entropy change with respect to the resting state ($V_{\mathrm{m}} = -70$ mV). Thus, based on the aforementioned definitions, heat will be released ($Q < 0$) from the membrane domain $\Omega_{\mathrm{M}}$ to the surroundings, as the (electric) internal energy of the membrane domain decreases ($\Delta{U_{\mathrm{el}}} < 0$). In the following sections, we will derive the free energy and entropy changes in the membrane domain as a function of the membrane potentials $V_{\mathrm{m}}$ and $\phi_{\mathrm{t}}$. The sum of these energies will give the internal energy change (by Eq. \eqref{eq:dUdef}) associated with the (de)polarization of the membrane and therefore the quantity of heat that is reversibly released from the membrane domain $\Omega_{\mathrm{M}}$.
\subsection{Electric free energy}\label{sec:free_energy}
\begin{figure}[h]
\includegraphics[width=1\linewidth]{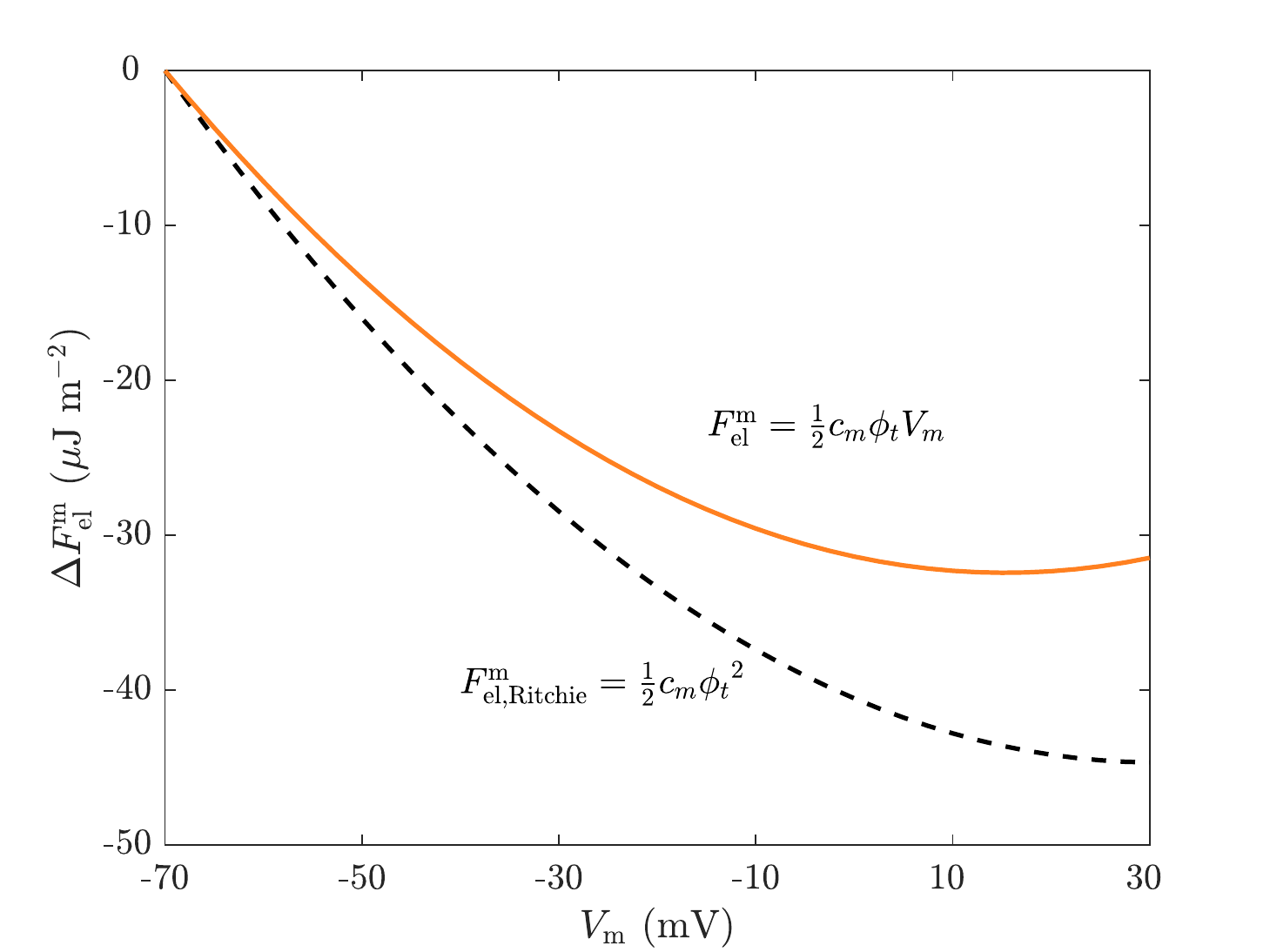}
\caption{Free energy changes in the membrane, according to the revised Condenser Theory (this work, $\Delta F^{\mathrm{m}}_{\mathrm{el}}$ ) and to the formula suggested by \citet{Howarth1979} and \citet{Ritchie1985} ($\Delta F^{\mathrm{m}}_{\mathrm{el,Ritchie}}$
). Strikingly, the latter overestimates free energy changes. Surface charge densities assumed are $\sigma_{\mathrm{i}} = -0.1$ C m$^{-2}$ and $\sigma_{\mathrm{o}} = -0.05 $ C m$^{-2}$.}
\label{fig:deltaF_ritchie}
\end{figure}

The electric free energy of a linear dielectric medium is equal to its field energy \cite{Overbeek1990,Jackson1999,landau2013electrodynamics}, that is, in one dimension
\begin{equation}\label{eq:free_el_energy}
F_{{\mathrm{el}}} = \frac{1}{2} \int^{+\infty}_{-\infty}{E.D}\:dx,
\end{equation}
where $F_{{\mathrm{el}}}$ is the electric free energy (J m$^{-2}$), $E$ the electric field (V m$^{-1}$) and $D = \varepsilon \: E$ the displacement field (C m$^{-2}$).
Interestingly, the field energy can be equated with the amount of heat dissipated by ionic currents in an electric field, using Maxwell's equation for Ampere's law (this is shown in Section   \ref{sec:Maxwell}). Note that the field energy is often regarded as an \textit{internal energy} ($U_{\mathrm{el}}$), however, this is only true in a primitive model that considers the medium as structureless \cite{Frohlich1968,Overbeek1990}. In a more refined model that takes into the effect of the electric field on the entropy of the dielectric medium (see Section\ref{sec:entropy}), the field energy must be regarded as a \textit{free energy} ($F_{\mathrm{el}}$) \cite{Frohlich1968,Overbeek1990}.
Applying Eq. \eqref{eq:free_el_energy} to the $\Omega_{\mathrm{M}}$ domain, we obtain the electric free energy as a function of the membrane potential $V_{\mathrm{m}}$. We obtain a free energy contribution from the double layers, which is
\begin{equation}\label{eq:FDL}
\begin{split}
{F}^{\mathrm{DL}}_{\mathrm{el}} =& \frac{1}{2}\bigg( \int_{-\infty}^{-\delta}{\rho_{\mathrm{i}}(\phi_{\mathrm{i}} - V_{\mathrm{m}})} dx + \sigma_{\mathrm{i}} \:(\phi_{\mathrm{i}}(-\delta) - V_{\mathrm{m}}) \\&+ \int^{+\infty}_{0}{\rho_{\mathrm{o}}\:\phi_{\mathrm{o}}\:}dx + \sigma_{\mathrm{o}}\:\phi_{\mathrm{o}}(0)\bigg),
\end{split}
\end{equation}
and one from the membrane capacitance, which is
\begin{equation}\label{eq:Fm}
{F}^{\mathrm{m}}_{\mathrm{el}} =  \frac{1}{2}\:c_{\mathrm{m}}\: \phi_{\mathrm{t}}\:V_{\mathrm{m}},
\end{equation}
with ${F}_{\mathrm{el}} = {F}^{\mathrm{DL}}_{\mathrm{el}} + {F}^{\mathrm{m}}_{\mathrm{el}}$. 
The first and third terms in Eq. \eqref{eq:FDL} correspond to the energy of bulk charges in the diffuse layers of internal and external solutions and the second and fourth to the energy of the surface charges fixed onto the membrane. As shown by Eq. \eqref{eq:Fm}, the free energy contribution of the membrane capacitance is $ {1}/{2}\:c_{\mathrm{m}}\:\phi_{\mathrm{t}}\:V_{\mathrm{m}}$, and not $ {1}/{2}c_{\mathrm{m}}\:\phi_{\mathrm{t}}^2$ as proposed in Refs. \cite{Howarth1979} and \cite{Ritchie1985}. The latter overestimates free energy changes, as compared to Eq. \eqref{eq:Fm}, see Fig. \ref{fig:deltaF_ritchie} in Section \ref{sec:suppl_plot}. The full derivation of Eqs. \eqref{eq:FDL} and \eqref{eq:Fm} is presented Section \ref{sec:free_energy_derivation}. 

Finally, we write the free energy \textit{changes} associated with the depolarisation of the membrane as $\Delta{F^{\mathrm{DL}}_{\mathrm{el}}} ={F^{\mathrm{DL}}_{\mathrm{el}}} - {F^{\mathrm{DL}}_{\mathrm{el,0}}}$ and $\Delta{F^{\mathrm{m}}_{\mathrm{el}}} ={F^{\mathrm{m}}_{\mathrm{el}}} - {F^{\mathrm{m}}_{\mathrm{el,0}}}$, where subscripts ``0'' mark the free energies calculated at the resting membrane potential, $V_{\mathrm{m,0}} = -70$ mV.
\subsection{Entropy associated with the electric field}\label{sec:entropy}
We consider two entropy terms: entropy changes associated with the polarization of water in the diffuse layers and entropy changes in the membrane.  
\subsubsection{In the diffuse layers}
The electric field orders water dipoles in the diffuse layers, which decreases entropy. The change of entropy associated with the alignment of dipoles in a dielectric medium (water in our case) is related to the electric free energy by \cite{Frohlich1968}:
\begin{equation}\label{eq:entropy-free-energy-DL}
T\Delta{S}^{\mathrm{DL}}_{\mathrm{el}} = \frac{T}{\varepsilon}\frac{\partial{\varepsilon}}{\partial{T}} \Delta{F^{\mathrm{DL}}_{\mathrm{el}}}
\end{equation}
where $\varepsilon$ is the dielectric permittivity of the medium. The value of ${T}/{\varepsilon}\:{\partial{\varepsilon}}/{\partial{T}}$ for water is -1.17 at 0\degree C and -1.4 at human body temperature (37\degree C) \cite{Floriano2004,lide2004crc}.\\ 
\subsubsection{In the membrane}
Similarly, entropy changes inside the lipid membrane have been proposed based on the temperature dependence of the membrane capacitance \cite{Howarth1968,Howarth1979,Ritchie1985}:
\begin{equation}\label{eq:entropy-free-energy-m}
T\Delta{S^{\mathrm{m}}_{\mathrm{el}}} = \frac{T}{c_{\mathrm{m}}}\:\frac{\partial{c_{\mathrm{m}}}}{\partial{T}}\:\Delta{F^{\mathrm{m}}_{\mathrm{el}}}.
\end{equation}
To verify that Eq. \eqref{eq:entropy-free-energy-m} holds, we adapted the derivation of Eq. \eqref{eq:entropy-free-energy-DL}, in which entropy is a function of ${\partial{\varepsilon}}/{\partial{T}}$, to the case in which entropy is a function of ${\partial{c_{\mathrm{m}}}}/{\partial{T}}$, the temperature dependence of the membrane capacitance. As shown in Section \ref{sec:entropy_derivation}, Eq. \eqref{eq:entropy-free-energy-m} holds, under the assumption that the membrane capacitance has a linear dependence on temperature, but is constant with potential.
Interestingly, it appears from recent experiments that the temperature dependence of the membrane capacitance arises from the variation of the lipid bilayer's dimensions (thickness, $\delta$, and area, $A$, per lipid molecule) with temperature, rather than from the one of its dielectric permittivity (${\partial{\varepsilon_{\mathrm{m}}}}/{\partial{T}}\simeq 0$) \cite{Plaksin2018,Kucerka2011,szekely2011effect,Pan2008}:
\begin{equation}
\frac{\partial{c_{\mathrm{m}}}}{\partial{T}} = \frac{\partial{(\varepsilon_{\mathrm{m}}\:A/\delta)}}{\partial{T}} \simeq \varepsilon_{\mathrm{m}} \frac{\partial{(A/\delta)}}{\partial{T}}.
\end{equation}
Based on recent measurements of these dimensional changes \cite{Kucerka2011,szekely2011effect,Pan2008} \citet{Plaksin2018} have pointed out that the temperature dependence of the membrane capacitance remains close to $+ 0.3 \%$/\degree C across several cellular types and artificial lipid membranes, suggesting that the rate of thermal response of the membrane is universal. 
\subsection{Parameters}
Two important parameters in this model are the surface charge density on the interior ($\sigma_{\mathrm{i}}$ ) and exterior ($\sigma_{\mathrm{o}}$) sides of the membrane. Here we report ranges of values found in literature and then choose baselines for these two parameters. \citet{Hille2001} compiled experimental data on excitable membranes, showing that the surface charge density (extracellular and intracellular) varies from $-0.04$ to $-0.16$ C m$^{-2}$. Other estimates give a wider range of values, from $-0.002$ to $-0.37$ C m$^{-2}$ \cite{Lakshminarayanaiah1975}.
In this work, the external surface charge density will be fixed to $\sigma_{\mathrm{o}} = -0.05$ C m$^{-2}$, and we will evaluate three cases for the internal surface charge density: $\sigma_{\mathrm{i}} = -0.05$, $-0.1$ and $-0.15$ C m$^{-2}$.
There are several lines of evidence suggesting that neurons have more negative surface charges on the internal side of the membrane than on the external one. 
In rat cortical neurons for instance, \citet{Plaksin2018} estimated this surface charge \textit{bias} to be $\sigma_{\mathrm{i}} - \sigma_{\mathrm{o}} =  - 0.1$ C m$^{-2}$.
Further support for this hypothesis arises from the uneven distribution of phospholipids between the two sides of the membrane. In particular, phosphatidylserine, the most abundant negatively charged phospholipid in cell membranes, is found exclusively on the internal side of the cell membrane of neurons, where it has key signaling functions \cite{kim2014phosphatidylserine}. 
Other parameters used in the model are reported in Table \ref{tab:parameters_heat}. 
\begin{table}[H]
\centering
\begin{tabular}{l l l l}
\toprule
Parameter & Value & Unit & Source  \\
\hline
$\delta$   & 3      & nm  &  \cite{Genet2000} \\
$c_{\mathrm{m}}$ & 9      & mF m$^{-2}$  &  \cite{Genet2000} \\
$V_{\mathrm{m}}$ & -70      & mV  &  \cite{weiss1996cellular,Hille2001,Genet2000} \\
${1}/{\varepsilon}\:{\partial{\varepsilon}}/{\partial{T}}$ & - 0.43 at 0\degree C &$\%$K$^{-1}$ &\cite{Floriano2004,lide2004crc}\\
${1}/{c_{\mathrm{m}}}\:{\partial{c_{\mathrm{m}}}}/{\partial{T}} $ &+ 0.3 &$\%$K$^{-1}$ &\cite{Plaksin2018}\\
$c_{\ce{Na+}}$ &5 / 145   & mM &\cite{johnston1994foundations}\\
$c_{\ce{K+}}$ &140 / 5 &mM&\cite{johnston1994foundations}\\
$c_{\ce{Ca^2+}}$ &0.0001 / 2.5 & mM&\cite{johnston1994foundations}\\
$c_{\ce{Cl-}}$ & 145 / 155 &mM&\cite{johnston1994foundations}\\
$T$    & 273 &K  & \cite{Howarth1979} \\
\bottomrule
\end{tabular}
\caption{Parameters used in the electrostatic model and energy calculations. Concentrations are given in the format: internal solution / external solution.}\label{tab:parameters_heat}  
\end{table}
\newpage
\section{Results}
\subsection{Free energy changes}
\begin{figure}[h]
\centering
\includegraphics[width=1\linewidth]{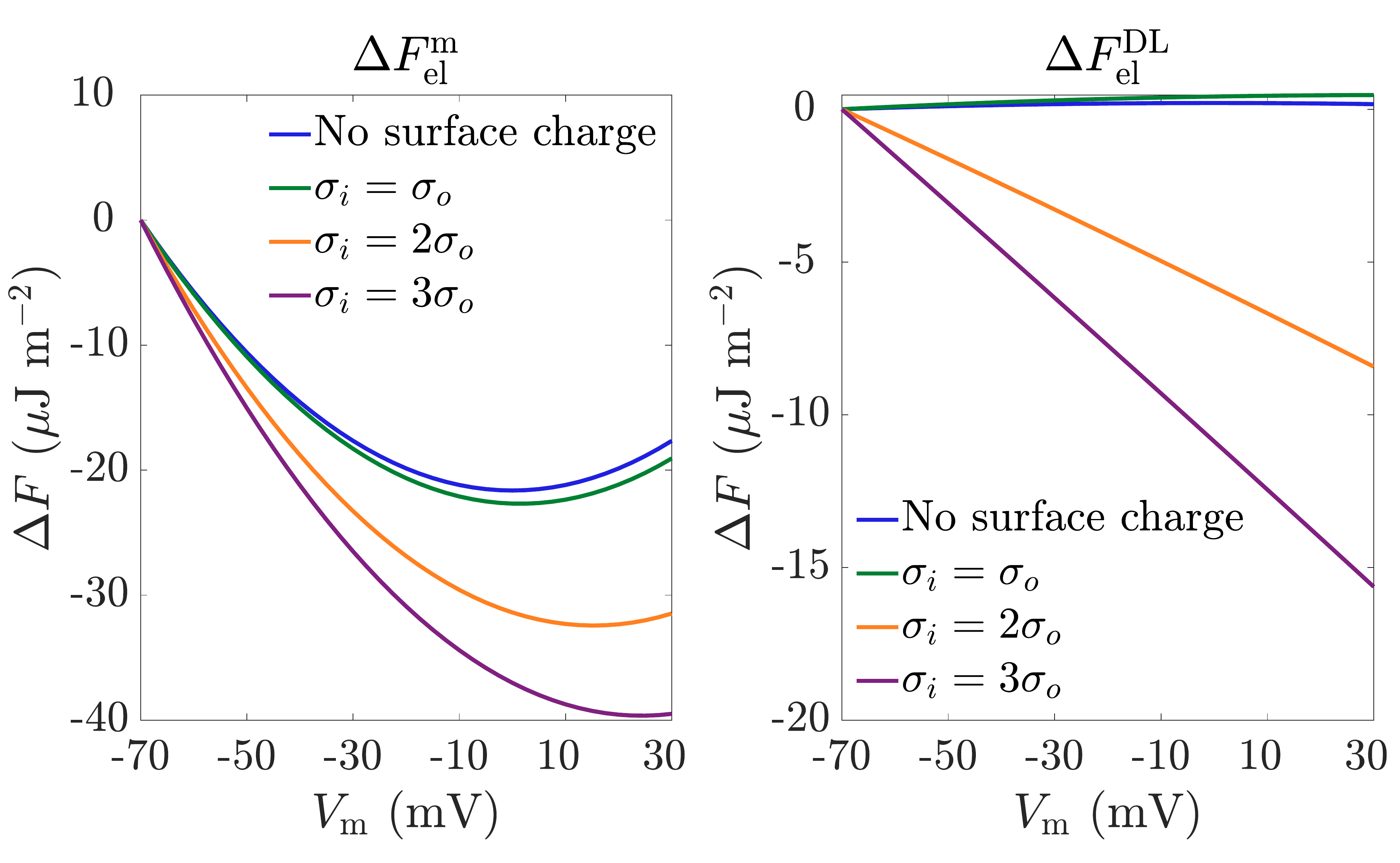}
\caption{Change in electric free energy in the membrane ($\Delta{F}^{\mathrm{m}}_{\mathrm{el}}$) and in diffuse layers ($\Delta{F}^{\mathrm{DL}}_{\mathrm{el}}$) as a function of membrane potential, for an equal amount of surface charges ($\sigma_{\mathrm{i}} =  \sigma_{\mathrm{o}} = -0.05$ C m$^{-2}$) and with a surface charge bias ($\sigma_{\mathrm{i}} = 2\:\sigma_{\mathrm{o}}$ and $ 3\:\sigma_{\mathrm{o}}$). When a surface charge bias is present, more free energy is released, both in the membrane and diffuse layers.}
\label{fig:delta_F}
\end{figure}

The change in electric free energy with membrane potential is depicted in Fig. \ref{fig:delta_F}, in the membrane ($\Delta{F^{\mathrm{m}}_{\mathrm{el}}}$) and in diffuse layers ($\Delta{F^{\mathrm{DL}}_{\mathrm{el}}}$), with and without surface charges on the membrane. First, when there are no surface charges, the free energy follows a parabola centred around $V_{\mathrm{m}}$ = 0 mV. The diffuse layers bring a relatively negligible contribution to the free energy in this case. Interestingly, the presence of an equal amount of surface charges on each side of the membrane ($\sigma_{\mathrm{i}} = \sigma_{\mathrm{o}} = - 0.05 $ C m$^{-2}$) results in almost no alteration of the energy changes as compared to the zero surface charge case.
However, as more surface charges are present on the internal side than on the external side of the membrane ($\sigma_{\mathrm{i}} = 2\:\sigma_{\mathrm{o}}$ and $3\:\sigma_{\mathrm{o}}$ ), both $\Delta{F^{\mathrm{m}}_{\mathrm{el}}}$ and $\Delta{F^{\mathrm{DL}}_{\mathrm{el}}}$ follow a steeper decrease, such that more free energy is released with increasing $V_{\mathrm{m}}$.

\subsection{Entropy changes}
\begin{figure}[h]
\centering
\includegraphics[width=1\linewidth]{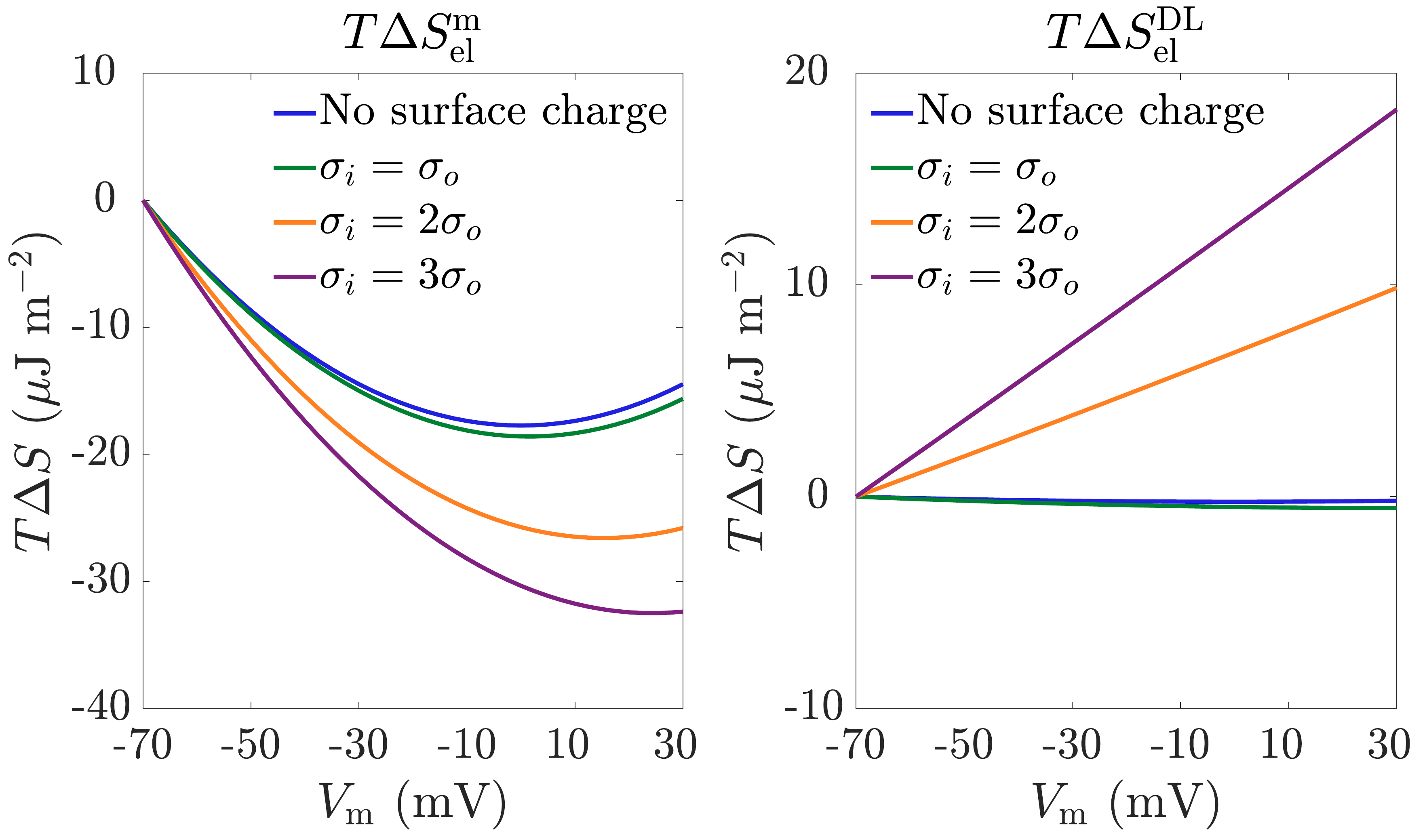}
\caption{Entropy changes in the membrane ($T\Delta{S}^{\mathrm{m}}_{\mathrm{el}}$) and diffuse layers ($T\Delta{S}^{\mathrm{DL}}_{\mathrm{el}}$) with no surface charges, an equal amount of surface charges ($\sigma_{\mathrm{i}} =  \sigma_{\mathrm{o}} = $-0.05 C m$^{-2}$) and with a surface charge bias ($\sigma_{\mathrm{i}} = 2\:\sigma_{\mathrm{o}}$ and $3\:\sigma_{\mathrm{o}}$). The presence of a surface charge bias makes the entropy in the membrane decrease more upon depolarization, whereas it makes the entropy in diffuse layers increase.}
\label{fig:delta_S}
\end{figure}

Entropy changes, depicted in Fig. \ref{fig:delta_S}, are proportional to the free energy changes presented above (by Eqs. \eqref{eq:entropy-free-energy-DL} and \eqref{eq:entropy-free-energy-m}). As the membrane depolarizes (increasing $V_{\mathrm{m}}$), the entropy in the membrane ($T\Delta{S}^{\mathrm{m}}_{\mathrm{el}}$) decreases, while the one in diffuse layers ($T\Delta{S}^{\mathrm{DL}}_{\mathrm{el}}$) increases. Both entropy changes follow the same trend with surface charge distribution as the free energy changes in Fig. \ref{fig:delta_F}. 

\subsection{Internal energy change, heat production and temperature change}\label{sec:internal_energy}
\begin{figure}[H]
\centering
\includegraphics[width=1\linewidth]{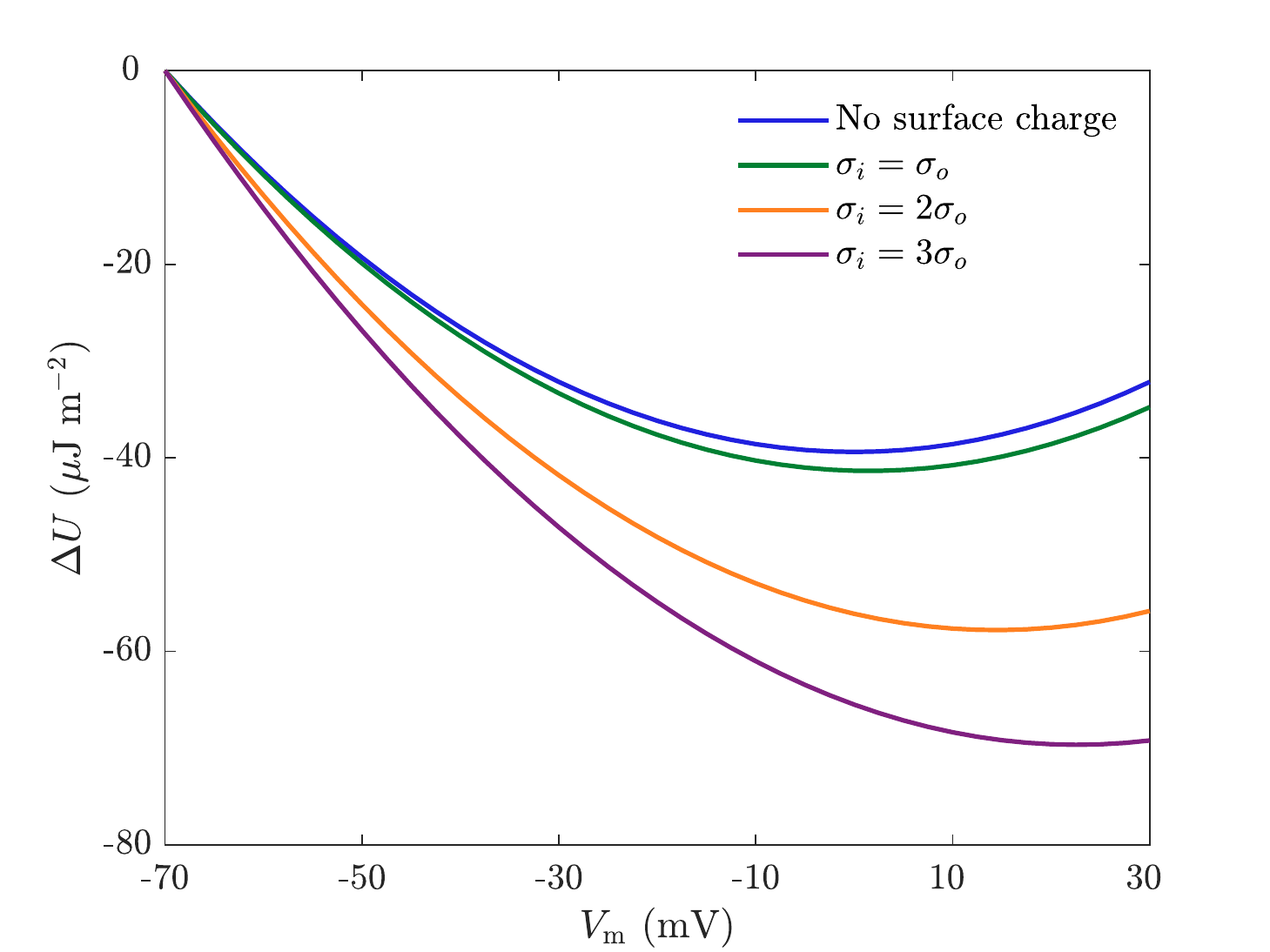}
\caption{Internal energy changes of the membrane system, from top to bottom: without surface charge ($\sigma_{\mathrm{i}} =  \sigma_{\mathrm{o}} = $ 0), with an equal amount of surface charges ($\sigma_{\mathrm{i}} =  \sigma_{\mathrm{o}} = -0.05 $ C m$^{-2}$), with a moderate surface charge bias ($\sigma_{\mathrm{i}} = 2\: \sigma_{\mathrm{o}}$) and a bigger one ($\sigma_{\mathrm{i}} = 3\:\sigma_{\mathrm{o}}$). The internal energy release increases with the presence of negative surface charges on each side of the membrane, and significantly more as the surface charge difference between the internal and external side of the membrane rises. }
\label{fig:internal_energy}
\end{figure}

\begin{figure}[H]
\centering
\includegraphics[width=1\linewidth]{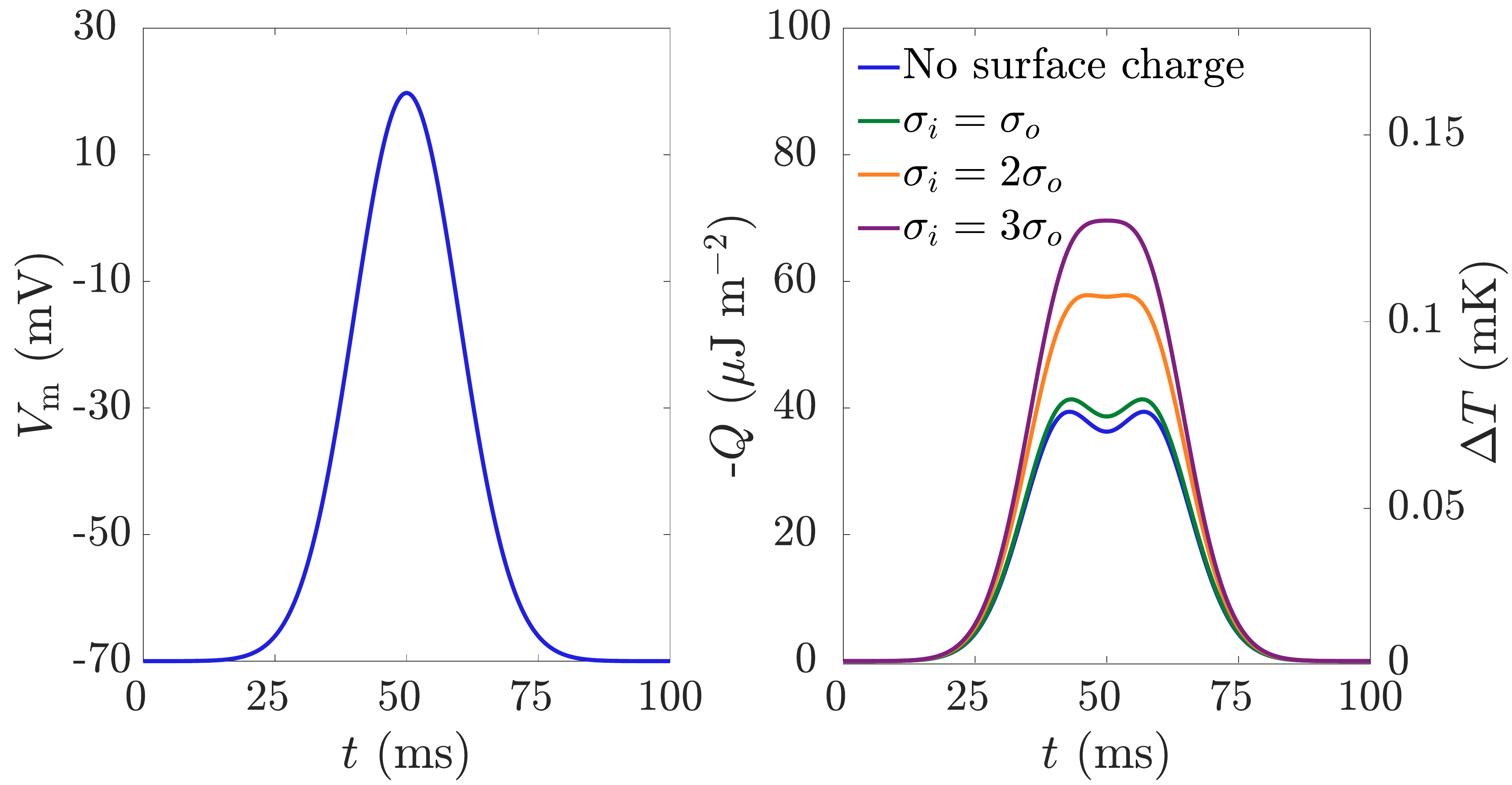}
\caption{Left: an action potential depolarizes the membrane, starting from a resting potential of $V_{\mathrm{m}}$ = -70 mV at time 0, the membrane depolarizes to  $V_{\mathrm{m}}$ = +20 mV at the peak of the action potential, before decreasing back to the resting potential. Right: concomitantly with the action potential, heat ($-Q$) is produced and absorbed, increasing the nerve temperature ($\Delta{T}$). See Section \ref{sec:fast-eq-T} for the calculation of temperature increase as a function of the heat released.}
\label{fig:VHT_t}
\end{figure}

The internal energy change occurring inside the $\Omega_{\mathrm{M}}$ domain, \textit{i.e} the sum of the free and the entropic energy changes, is depicted in Fig. \ref{fig:internal_energy} according to the different scenarios of surface charge distribution used until now. The internal energy change provides a direct measure of the heat produced and absorbed by the nerve during the course of the action potential (Eq. \eqref{eq:dU_Q}): an internal energy decrease ($\Delta{U}<0$) corresponds to a release of heat by the membrane domain ($Q<0$), whereas an internal energy increase ($\Delta{U}>0$) corresponds to an absorption of heat by the membrane domain ($Q>0$). An idealized action potential (modelled with a normal distribution function, for simplicity) and the corresponding heat profile are depicted in Fig. \ref{fig:VHT_t}. We find that the presence of negative surface charges on the membrane leads to a more important decrease in internal energy (Fig. \ref{fig:internal_energy}) and thus heat production (Fig. \ref{fig:VHT_t}), especially when the membrane holds more fixed charges on its inside than outside. Interestingly, the bottom curve in Fig. \ref{fig:internal_energy} shows that when surface charges are distributed unevenly, the internal energy must not rise immediately after that $V_{\mathrm{m}}$ takes positive values. In other words, the membrane can release heat even when the membrane potential overshoots to positive values. This is reflected in Fig. \ref{fig:VHT_t} by the progressive disappearance of the ``notch'' in the energy profiles as $\sigma_{\mathrm{i}}$ becomes more negative than $\sigma_{\mathrm{o}}$. 

Finally, based on Fig. \ref{fig:VHT_t}, we predict that a typical depolarization of the membrane from -70 mV to +20 mV \cite{weiss1996cellular,dudel2012fundamentals} leads to a heat production of $-Q = 40$ to \SI{70}{\micro\joule} m$^{-2}$, depending on the magnitude of the bias of surface charges present on the membrane (0 to -0.1 C m$^{-2}$). If the action potential starts from a lower resting potential (which can be down to -100 mV \cite{Hille2001}), up to \SI{150}{\micro\joule} m$^{-2}$ of heat is predicted by the model. The predicted values fall in the range of experimental values (60 - \SI{180}{\micro\joule} m$^{-2}$) \cite{Howarth1979}).
\section{Discussion}
The purpose of this work was to provide a theoretical background for heat production and absorption in neurons that goes beyond the simplistic analogy with a capacitor. Based on an equilibrium thermodynamics description of the membrane, its surface charges and double layers forming on each side, we evaluated the amount of heat released due to free energy and entropy changes during the action potential. The model assumes complete reversibility: all internal energy changes are first converted into heat, and then all heat is converted back to a change in the internal energy. 

First, we found that the electric free energy of the cell membrane depends on both the membrane potential and the transmembrane potential, as shown by the new expression derived in this work: ${F}^{\mathrm{m}}_{\mathrm{el}} = \frac{1}{2}c_{\mathrm{m}}\:\phi_{\mathrm{t}}\:V_{\mathrm{m}}$. In consequence, the amount of free energy stored by the membrane increases as more surface charges are present on the internal versus the external side of the membrane. 
Free energy changes in the diffuse layers surrounding the membrane also increase significantly when a bias of surface charge is present, however, these changes are offset by the entropy changes associated with the polarization of water by the electric field, that are of comparable magnitude but of opposite sign ($\sim$-120\% at 0\degree C).
Finally, entropy changes in the membrane appear to be related to the dependence on temperature of the lipid bilayer's dimensions rather than of its dielectric permittivity. Recent measurements of the dimensional variations allowed to determine that the magnitude of entropy changes is comparable ($\sim$+82$\%$) to the one of free energy changes.

The comparison of model predictions with the best-available measurements of the heat production and absorption in neurons supports the idea that the heat of nervous conduction has an electrical origin: %
 assuming a typical action potential that depolarizes the membrane from -70 mV to +20 mV and a surface charge bias of $-0.05$ C m$^{-2}$ between the two sides of the membrane, we predict a production and absorption of heat of $\sim$60 \SI{}{\micro\joule} m$^{-2}$, which is sufficient to reach the range of experimental values measured in the neurons of different organisms (60 - 180 \SI{}{\micro\joule} m$^{-2}$) \cite{Howarth1979}. Should the size of the action potential be larger (starting from a more negative resting potential), or the surface charge bias be more pronounced, more heat would be evolved.
 
Clearly, our predictions have a critical dependency on the action potential size and surface charge distribution between the internal and external sides of the membrane. Both parameters can vary from one type of neuron to the other, however, the values used for our calculations do not seem extravagant, considering the range of values reported in literature for surface charge density \cite{Hille2001,Lakshminarayanaiah1975}.
The experimental validation of our predictions requires to compare the heat production with the true course of the action potential in neurons, which can only be obtained by measuring the membrane potential intracellularly. With the recent development of nano-electrodes \cite{Angle2015,Duan2013,Ferguson2012}, there is hope that it is becoming possible to record true membrane potentials in the small neurons in which temperature changes can effectively be measured.
Although we cannot discard the possibility of other sources of heat, such as heat arising from electrically-induced elastic displacements of the membrane \cite{ElHady2015,Lacoste2009,Lacoste2007}, our analysis suggests that the change in electrical energy of the membrane represents the most prominent mechanism of heat production and absorption by neurons during the propagation of the action potential.

Besides neurophysiology, the physics our work explores could be relevant for engineering electrical double layer capacitors \cite{janssen2017reversible} or designing capacitive deionization processes, which performance can be influenced by the presence of surface charges, either immobile \cite{Biesheuvel2015} or modulated electrochemically \cite{he2018theory}. Moreover, thermodynamic analyses in capacitive systems typically do not include the entropy of dipoles in dielectric media \cite{hemmatifar2018thermodynamics}. Here, we showed that the entropy of dipoles brings a significant contribution to the internal energy stored across the double layers and the associated heat production and absorption. 

\section{Author Information}
\noindent Corresponding Authors:
\noindent$^*$E-mail: bazant@mit.edu \\
\noindent Author Contributions:\\
$^\dagger$ A.D.L and J.P.D contributed equally to this work.
\section{Acknowledgements}
J.P.D. acknowledges support in the form of a National Science Foundation Graduate Research Fellowship under Grant No. 1122374.
A.D.L. acknowledges support from the Louis-de-Broglie Foundation of the French Academy of Sciences. The authors thank Mathijs Janssen and Maarten Biesheuvel for useful comments.
\bibliography{library}

\begin{thebibliography}{51}%
\makeatletter
\providecommand \@ifxundefined [1]{%
 \@ifx{#1\undefined}
}%
\providecommand \@ifnum [1]{%
 \ifnum #1\expandafter \@firstoftwo
 \else \expandafter \@secondoftwo
 \fi
}%
\providecommand \@ifx [1]{%
 \ifx #1\expandafter \@firstoftwo
 \else \expandafter \@secondoftwo
 \fi
}%
\providecommand \natexlab [1]{#1}%
\providecommand \enquote  [1]{``#1''}%
\providecommand \bibnamefont  [1]{#1}%
\providecommand \bibfnamefont [1]{#1}%
\providecommand \citenamefont [1]{#1}%
\providecommand \href@noop [0]{\@secondoftwo}%
\providecommand \href [0]{\begingroup \@sanitize@url \@href}%
\providecommand \@href[1]{\@@startlink{#1}\@@href}%
\providecommand \@@href[1]{\endgroup#1\@@endlink}%
\providecommand \@sanitize@url [0]{\catcode `\\12\catcode `\$12\catcode
  `\&12\catcode `\#12\catcode `\^12\catcode `\_12\catcode `\%12\relax}%
\providecommand \@@startlink[1]{}%
\providecommand \@@endlink[0]{}%
\providecommand \url  [0]{\begingroup\@sanitize@url \@url }%
\providecommand \@url [1]{\endgroup\@href {#1}{\urlprefix }}%
\providecommand \urlprefix  [0]{URL }%
\providecommand \Eprint [0]{\href }%
\providecommand \doibase [0]{http://dx.doi.org/}%
\providecommand \selectlanguage [0]{\@gobble}%
\providecommand \bibinfo  [0]{\@secondoftwo}%
\providecommand \bibfield  [0]{\@secondoftwo}%
\providecommand \translation [1]{[#1]}%
\providecommand \BibitemOpen [0]{}%
\providecommand \bibitemStop [0]{}%
\providecommand \bibitemNoStop [0]{.\EOS\space}%
\providecommand \EOS [0]{\spacefactor3000\relax}%
\providecommand \BibitemShut  [1]{\csname bibitem#1\endcsname}%
\let\auto@bib@innerbib\@empty
\bibitem [{\citenamefont {Hodgkin}\ \emph {et~al.}(1952)\citenamefont
  {Hodgkin}, \citenamefont {Huxley},\ and\ \citenamefont
  {Katz}}]{hodgkin1952measurement}%
  \BibitemOpen
  \bibfield  {author} {\bibinfo {author} {\bibfnamefont {A.~L.}\ \bibnamefont
  {Hodgkin}}, \bibinfo {author} {\bibfnamefont {A.~F.}\ \bibnamefont {Huxley}},
  \ and\ \bibinfo {author} {\bibfnamefont {B.}~\bibnamefont {Katz}},\
  }\href@noop {} {\bibfield  {journal} {\bibinfo  {journal} {The Journal of
  physiology}\ }\textbf {\bibinfo {volume} {116}},\ \bibinfo {pages} {424}
  (\bibinfo {year} {1952})}\BibitemShut {NoStop}%
\bibitem [{\citenamefont {Hodgkin}\ and\ \citenamefont
  {Huxley}(1952)}]{hodgkin1952components}%
  \BibitemOpen
  \bibfield  {author} {\bibinfo {author} {\bibfnamefont {A.~L.}\ \bibnamefont
  {Hodgkin}}\ and\ \bibinfo {author} {\bibfnamefont {A.~F.}\ \bibnamefont
  {Huxley}},\ }\href@noop {} {\bibfield  {journal} {\bibinfo  {journal} {The
  Journal of physiology}\ }\textbf {\bibinfo {volume} {116}},\ \bibinfo {pages}
  {473} (\bibinfo {year} {1952})}\BibitemShut {NoStop}%
\bibitem [{\citenamefont {Neher}\ and\ \citenamefont
  {Sakmann}(1992)}]{neher1992}%
  \BibitemOpen
  \bibfield  {author} {\bibinfo {author} {\bibfnamefont {E.}~\bibnamefont
  {Neher}}\ and\ \bibinfo {author} {\bibfnamefont {B.}~\bibnamefont
  {Sakmann}},\ }\href@noop {} {\bibfield  {journal} {\bibinfo  {journal}
  {Scientific American}\ }\textbf {\bibinfo {volume} {266}},\ \bibinfo {pages}
  {44} (\bibinfo {year} {1992})}\BibitemShut {NoStop}%
\bibitem [{\citenamefont {Hamill}\ \emph {et~al.}(1981)\citenamefont {Hamill},
  \citenamefont {Marty}, \citenamefont {Neher}, \citenamefont {Sakmann},\ and\
  \citenamefont {Sigworth}}]{hamill1981}%
  \BibitemOpen
  \bibfield  {author} {\bibinfo {author} {\bibfnamefont {O.~P.}\ \bibnamefont
  {Hamill}}, \bibinfo {author} {\bibfnamefont {A.}~\bibnamefont {Marty}},
  \bibinfo {author} {\bibfnamefont {E.}~\bibnamefont {Neher}}, \bibinfo
  {author} {\bibfnamefont {B.}~\bibnamefont {Sakmann}}, \ and\ \bibinfo
  {author} {\bibfnamefont {F.}~\bibnamefont {Sigworth}},\ }\href@noop {}
  {\bibfield  {journal} {\bibinfo  {journal} {Pfl{\"u}gers Archiv}\ }\textbf
  {\bibinfo {volume} {391}},\ \bibinfo {pages} {85} (\bibinfo {year}
  {1981})}\BibitemShut {NoStop}%
\bibitem [{\citenamefont {Howarth}\ \emph {et~al.}(1979)\citenamefont
  {Howarth}, \citenamefont {Ritchie},\ and\ \citenamefont
  {Stagg}}]{Howarth1979}%
  \BibitemOpen
  \bibfield  {author} {\bibinfo {author} {\bibfnamefont {J.~V.}\ \bibnamefont
  {Howarth}}, \bibinfo {author} {\bibfnamefont {J.~M.}\ \bibnamefont
  {Ritchie}}, \ and\ \bibinfo {author} {\bibfnamefont {D.}~\bibnamefont
  {Stagg}},\ }\href@noop {} {\bibfield  {journal} {\bibinfo  {journal} {Proc.
  R. Soc. Lond.}\ }\textbf {\bibinfo {volume} {367}},\ \bibinfo {pages} {347 }
  (\bibinfo {year} {1979})}\BibitemShut {NoStop}%
\bibitem [{\citenamefont {Tasaki}\ \emph {et~al.}(1968)\citenamefont {Tasaki},
  \citenamefont {Watanabe}, \citenamefont {Sandlin},\ and\ \citenamefont
  {Carnay}}]{Tasaki1968}%
  \BibitemOpen
  \bibfield  {author} {\bibinfo {author} {\bibfnamefont {I.}~\bibnamefont
  {Tasaki}}, \bibinfo {author} {\bibfnamefont {A.}~\bibnamefont {Watanabe}},
  \bibinfo {author} {\bibfnamefont {R.}~\bibnamefont {Sandlin}}, \ and\
  \bibinfo {author} {\bibfnamefont {L.}~\bibnamefont {Carnay}},\ }\href
  {\doibase 10.1073/pnas.61.3.883} {\bibfield  {journal} {\bibinfo  {journal}
  {Proceedings of the National Academy of Sciences}\ }\textbf {\bibinfo
  {volume} {61}},\ \bibinfo {pages} {883} (\bibinfo {year} {1968})}\BibitemShut
  {NoStop}%
\bibitem [{\citenamefont {Tasaki}\ \emph {et~al.}(1989)\citenamefont {Tasaki},
  \citenamefont {Kusano},\ and\ \citenamefont {Byrne}}]{Tasaki1989}%
  \BibitemOpen
  \bibfield  {author} {\bibinfo {author} {\bibfnamefont {I.}~\bibnamefont
  {Tasaki}}, \bibinfo {author} {\bibfnamefont {K.}~\bibnamefont {Kusano}}, \
  and\ \bibinfo {author} {\bibfnamefont {P.~M.}\ \bibnamefont {Byrne}},\ }\href
  {\doibase 10.1016/S0006-3495(89)82902-9} {\bibfield  {journal} {\bibinfo
  {journal} {Biophysical Journal}\ }\textbf {\bibinfo {volume} {55}},\ \bibinfo
  {pages} {1033} (\bibinfo {year} {1989})}\BibitemShut {NoStop}%
\bibitem [{\citenamefont {Gonzalez-Perez}\ \emph {et~al.}(2016)\citenamefont
  {Gonzalez-Perez}, \citenamefont {Mosgaard}, \citenamefont {Budvytyte},
  \citenamefont {Villagran-Vargas}, \citenamefont {Jackson},\ and\
  \citenamefont {Heimburg}}]{Gonzalez-Perez2016}%
  \BibitemOpen
  \bibfield  {author} {\bibinfo {author} {\bibfnamefont {A.}~\bibnamefont
  {Gonzalez-Perez}}, \bibinfo {author} {\bibfnamefont {L.~D.}\ \bibnamefont
  {Mosgaard}}, \bibinfo {author} {\bibfnamefont {R.}~\bibnamefont {Budvytyte}},
  \bibinfo {author} {\bibfnamefont {E.}~\bibnamefont {Villagran-Vargas}},
  \bibinfo {author} {\bibfnamefont {A.~D.}\ \bibnamefont {Jackson}}, \ and\
  \bibinfo {author} {\bibfnamefont {T.}~\bibnamefont {Heimburg}},\ }\href
  {\doibase 10.1016/j.bpc.2016.06.005} {\bibfield  {journal} {\bibinfo
  {journal} {Biophysical Chemistry}\ }\textbf {\bibinfo {volume} {216}},\
  \bibinfo {pages} {51} (\bibinfo {year} {2016})}\BibitemShut {NoStop}%
\bibitem [{\citenamefont {Hille}(2001)}]{Hille2001}%
  \BibitemOpen
  \bibfield  {author} {\bibinfo {author} {\bibfnamefont {B.}~\bibnamefont
  {Hille}},\ }\href@noop {} {\emph {\bibinfo {title} {Ion channels of excitable
  membranes}}},\ Vol.\ \bibinfo {volume} {507}\ (\bibinfo  {publisher} {Sinauer
  Sunderland, MA},\ \bibinfo {year} {2001})\BibitemShut {NoStop}%
\bibitem [{\citenamefont {{El Hady}}\ and\ \citenamefont
  {Machta}(2015)}]{ElHady2015}%
  \BibitemOpen
  \bibfield  {author} {\bibinfo {author} {\bibfnamefont {A.}~\bibnamefont {{El
  Hady}}}\ and\ \bibinfo {author} {\bibfnamefont {B.~B.}\ \bibnamefont
  {Machta}},\ }\href {\doibase 10.1038/ncomms7697} {\bibfield  {journal}
  {\bibinfo  {journal} {Nature Communications}\ }\textbf {\bibinfo {volume}
  {6}},\ \bibinfo {pages} {6697} (\bibinfo {year} {2015})}\BibitemShut
  {NoStop}%
\bibitem [{\citenamefont {Abbott}\ \emph {et~al.}(1958)\citenamefont {Abbott},
  \citenamefont {Hill},\ and\ \citenamefont {Howarth}}]{Abbott1958}%
  \BibitemOpen
  \bibfield  {author} {\bibinfo {author} {\bibfnamefont {B.~C.}\ \bibnamefont
  {Abbott}}, \bibinfo {author} {\bibfnamefont {A.~V.}\ \bibnamefont {Hill}}, \
  and\ \bibinfo {author} {\bibfnamefont {J.~V.}\ \bibnamefont {Howarth}},\
  }\href {\doibase 10.1098/rspb.1958.0012} {\bibfield  {journal} {\bibinfo
  {journal} {Proceedings of the Royal Society B: Biological Sciences}\ }\textbf
  {\bibinfo {volume} {148}},\ \bibinfo {pages} {149} (\bibinfo {year}
  {1958})}\BibitemShut {NoStop}%
\bibitem [{\citenamefont {Howarth}\ and\ \citenamefont
  {Ritchie}(1968)}]{Howarth1968}%
  \BibitemOpen
  \bibfield  {author} {\bibinfo {author} {\bibfnamefont {K.~R.~D.}\
  \bibnamefont {Howarth}, \bibfnamefont {J.}}\ and\ \bibinfo {author}
  {\bibfnamefont {J.~M.}\ \bibnamefont {Ritchie}},\ }\href@noop {} {\bibfield
  {journal} {\bibinfo  {journal} {The Journal of Physiology}\ ,\ \bibinfo
  {pages} {745 }} (\bibinfo {year} {1968})}\BibitemShut {NoStop}%
\bibitem [{\citenamefont {Ritchie}\ and\ \citenamefont
  {Keynes}(1985)}]{Ritchie1985}%
  \BibitemOpen
  \bibfield  {author} {\bibinfo {author} {\bibfnamefont {J.~M.}\ \bibnamefont
  {Ritchie}}\ and\ \bibinfo {author} {\bibfnamefont {R.~D.}\ \bibnamefont
  {Keynes}},\ }\href@noop {} {\bibfield  {journal} {\bibinfo  {journal}
  {Quaterly Review of Biophysics}\ }\textbf {\bibinfo {volume} {18}},\ \bibinfo
  {pages} {451 } (\bibinfo {year} {1985})}\BibitemShut {NoStop}%
\bibitem [{\citenamefont {Ritchie}(1973)}]{Ritchie1973}%
  \BibitemOpen
  \bibfield  {author} {\bibinfo {author} {\bibfnamefont {J.}~\bibnamefont
  {Ritchie}},\ }\href@noop {} {\bibfield  {journal} {\bibinfo  {journal}
  {Progress in biophysics and molecular biology}\ }\textbf {\bibinfo {volume}
  {26}},\ \bibinfo {pages} {147} (\bibinfo {year} {1973})}\BibitemShut
  {NoStop}%
\bibitem [{\citenamefont {Easton}(1971)}]{easton1971}%
  \BibitemOpen
  \bibfield  {author} {\bibinfo {author} {\bibfnamefont {D.~M.}\ \bibnamefont
  {Easton}},\ }\href@noop {} {\bibfield  {journal} {\bibinfo  {journal}
  {Science}\ }\textbf {\bibinfo {volume} {172}},\ \bibinfo {pages} {952}
  (\bibinfo {year} {1971})}\BibitemShut {NoStop}%
\bibitem [{\citenamefont {Weiss}(1996)}]{weiss1996cellular}%
  \BibitemOpen
  \bibfield  {author} {\bibinfo {author} {\bibfnamefont {T.~F.}\ \bibnamefont
  {Weiss}},\ }\href@noop {} {\emph {\bibinfo {title} {Cellular biophysics}}},\
  Vol.~\bibinfo {volume} {1}\ (\bibinfo  {publisher} {MIT press Cambridge,
  Massachusetts},\ \bibinfo {year} {1996})\BibitemShut {NoStop}%
\bibitem [{\citenamefont {Taylor}\ \emph {et~al.}(1962)\citenamefont {Taylor},
  \citenamefont {Chandler},\ and\ \citenamefont {Knox}}]{taylor1962}%
  \BibitemOpen
  \bibfield  {author} {\bibinfo {author} {\bibfnamefont {R.}~\bibnamefont
  {Taylor}}, \bibinfo {author} {\bibfnamefont {W.}~\bibnamefont {Chandler}}, \
  and\ \bibinfo {author} {\bibfnamefont {W.}~\bibnamefont {Knox}},\ }\href@noop
  {} {\bibfield  {journal} {\bibinfo  {journal} {Biophysical Society
  Abstracts}\ } (\bibinfo {year} {1962})}\BibitemShut {NoStop}%
\bibitem [{\citenamefont {Plaksin}\ \emph {et~al.}(2018)\citenamefont
  {Plaksin}, \citenamefont {Shapira}, \citenamefont {Kimmel},\ and\
  \citenamefont {Shoham}}]{Plaksin2018}%
  \BibitemOpen
  \bibfield  {author} {\bibinfo {author} {\bibfnamefont {M.}~\bibnamefont
  {Plaksin}}, \bibinfo {author} {\bibfnamefont {E.}~\bibnamefont {Shapira}},
  \bibinfo {author} {\bibfnamefont {E.}~\bibnamefont {Kimmel}}, \ and\ \bibinfo
  {author} {\bibfnamefont {S.}~\bibnamefont {Shoham}},\ }\href {\doibase
  10.1103/PhysRevX.8.011043} {\bibfield  {journal} {\bibinfo  {journal}
  {Physical Review X}\ }\textbf {\bibinfo {volume} {8}},\ \bibinfo {pages}
  {11043} (\bibinfo {year} {2018})}\BibitemShut {NoStop}%
\bibitem [{\citenamefont {Lizhi}\ \emph {et~al.}(2008)\citenamefont {Lizhi},
  \citenamefont {Toyoda},\ and\ \citenamefont {Ihara}}]{Lizhi2008}%
  \BibitemOpen
  \bibfield  {author} {\bibinfo {author} {\bibfnamefont {H.}~\bibnamefont
  {Lizhi}}, \bibinfo {author} {\bibfnamefont {K.}~\bibnamefont {Toyoda}}, \
  and\ \bibinfo {author} {\bibfnamefont {I.}~\bibnamefont {Ihara}},\ }\href
  {\doibase 10.1016/j.jfoodeng.2007.12.035} {\bibfield  {journal} {\bibinfo
  {journal} {Journal of Food Engineering}\ }\textbf {\bibinfo {volume} {88}},\
  \bibinfo {pages} {151} (\bibinfo {year} {2008})}\BibitemShut {NoStop}%
\bibitem [{\citenamefont {Genet}\ \emph {et~al.}(2000)\citenamefont {Genet},
  \citenamefont {Costalat},\ and\ \citenamefont {Burger}}]{Genet2000}%
  \BibitemOpen
  \bibfield  {author} {\bibinfo {author} {\bibfnamefont {S.}~\bibnamefont
  {Genet}}, \bibinfo {author} {\bibfnamefont {R.}~\bibnamefont {Costalat}}, \
  and\ \bibinfo {author} {\bibfnamefont {J.}~\bibnamefont {Burger}},\
  }\href@noop {} {\bibfield  {journal} {\bibinfo  {journal} {Acta
  Biotheoretica}\ }\textbf {\bibinfo {volume} {48}},\ \bibinfo {pages} {273}
  (\bibinfo {year} {2000})}\BibitemShut {NoStop}%
\bibitem [{\citenamefont {Overbeek}(1990)}]{Overbeek1990}%
  \BibitemOpen
  \bibfield  {author} {\bibinfo {author} {\bibfnamefont {J.~T.~G.}\
  \bibnamefont {Overbeek}},\ }\href@noop {} {\bibfield  {journal} {\bibinfo
  {journal} {Colloids and Surfaces}\ }\textbf {\bibinfo {volume} {51}},\
  \bibinfo {pages} {61} (\bibinfo {year} {1990})}\BibitemShut {NoStop}%
\bibitem [{\citenamefont {Jackson}(1999)}]{Jackson1999}%
  \BibitemOpen
  \bibfield  {author} {\bibinfo {author} {\bibfnamefont {J.~D.}\ \bibnamefont
  {Jackson}},\ }\href@noop {} {\emph {\bibinfo {title} {Classical
  electrodynamics}}}\ (\bibinfo  {publisher} {AAPT},\ \bibinfo {year}
  {1999})\BibitemShut {NoStop}%
\bibitem [{\citenamefont {Landau}\ \emph {et~al.}(2013)\citenamefont {Landau},
  \citenamefont {Bell}, \citenamefont {Kearsley}, \citenamefont {Pitaevskii},
  \citenamefont {Lifshitz},\ and\ \citenamefont
  {Sykes}}]{landau2013electrodynamics}%
  \BibitemOpen
  \bibfield  {author} {\bibinfo {author} {\bibfnamefont {L.~D.}\ \bibnamefont
  {Landau}}, \bibinfo {author} {\bibfnamefont {J.}~\bibnamefont {Bell}},
  \bibinfo {author} {\bibfnamefont {M.}~\bibnamefont {Kearsley}}, \bibinfo
  {author} {\bibfnamefont {L.}~\bibnamefont {Pitaevskii}}, \bibinfo {author}
  {\bibfnamefont {E.}~\bibnamefont {Lifshitz}}, \ and\ \bibinfo {author}
  {\bibfnamefont {J.}~\bibnamefont {Sykes}},\ }\href@noop {} {\emph {\bibinfo
  {title} {Electrodynamics of continuous media}}},\ Vol.~\bibinfo {volume} {8}\
  (\bibinfo  {publisher} {Elsevier},\ \bibinfo {year} {2013})\BibitemShut
  {NoStop}%
\bibitem [{\citenamefont {Frohlich}(1968)}]{Frohlich1968}%
  \BibitemOpen
  \bibfield  {author} {\bibinfo {author} {\bibfnamefont {H.}~\bibnamefont
  {Frohlich}},\ }\href {https://books.google.be/books?id=qPnStAEACAAJ} {\emph
  {\bibinfo {title} {Theory of Dielectrics: Dielectric Constant and Dielectric
  Loss}}},\ Monographs on the physics and chemistry of materials\ (\bibinfo
  {publisher} {Oxford University Press},\ \bibinfo {year} {1968})\BibitemShut
  {NoStop}%
\bibitem [{\citenamefont {Floriano}\ and\ \citenamefont
  {Nascimento}(2004)}]{Floriano2004}%
  \BibitemOpen
  \bibfield  {author} {\bibinfo {author} {\bibfnamefont {W.~B.}\ \bibnamefont
  {Floriano}}\ and\ \bibinfo {author} {\bibfnamefont {M.~A.~C.}\ \bibnamefont
  {Nascimento}},\ }\href
  {http://www.scielo.br/scielo.php?script=sci_arttext&pid=S0103-97332004000100006&nrm=iso}
  {\bibfield  {journal} {\bibinfo  {journal} {{Brazilian Journal of Physics}}\
  }\textbf {\bibinfo {volume} {34}},\ \bibinfo {pages} {38 } (\bibinfo {year}
  {2004})}\BibitemShut {NoStop}%
\bibitem [{\citenamefont {Lide}(2004)}]{lide2004crc}%
  \BibitemOpen
  \bibfield  {author} {\bibinfo {author} {\bibfnamefont {D.~R.}\ \bibnamefont
  {Lide}},\ }\href@noop {} {\emph {\bibinfo {title} {CRC handbook of chemistry
  and physics}}},\ Vol.~\bibinfo {volume} {85}\ (\bibinfo  {publisher} {CRC
  press},\ \bibinfo {year} {2004})\BibitemShut {NoStop}%
\bibitem [{\citenamefont {Norbert~Kucerka}\ and\ \citenamefont
  {Katsaras}(2011)}]{Kucerka2011}%
  \BibitemOpen
  \bibfield  {author} {\bibinfo {author} {\bibfnamefont {M.-P.~N.}\
  \bibnamefont {Norbert~Kucerka}}\ and\ \bibinfo {author} {\bibfnamefont
  {J.}~\bibnamefont {Katsaras}},\ }\href {\doibase
  https://doi.org/10.1016/j.bbamem.2011.07.022} {\bibfield  {journal} {\bibinfo
   {journal} {Biochimica et Biophysica Acta (BBA) - Biomembranes}\ }\textbf
  {\bibinfo {volume} {1808}},\ \bibinfo {pages} {2761 } (\bibinfo {year}
  {2011})}\BibitemShut {NoStop}%
\bibitem [{\citenamefont {Szekely}\ \emph {et~al.}(2011)\citenamefont
  {Szekely}, \citenamefont {Dvir}, \citenamefont {Asor}, \citenamefont {Resh},
  \citenamefont {Steiner}, \citenamefont {Szekely}, \citenamefont {Ginsburg},
  \citenamefont {Mosenkis}, \citenamefont {Guralnick}, \citenamefont {Dan}
  \emph {et~al.}}]{szekely2011effect}%
  \BibitemOpen
  \bibfield  {author} {\bibinfo {author} {\bibfnamefont {P.}~\bibnamefont
  {Szekely}}, \bibinfo {author} {\bibfnamefont {T.}~\bibnamefont {Dvir}},
  \bibinfo {author} {\bibfnamefont {R.}~\bibnamefont {Asor}}, \bibinfo {author}
  {\bibfnamefont {R.}~\bibnamefont {Resh}}, \bibinfo {author} {\bibfnamefont
  {A.}~\bibnamefont {Steiner}}, \bibinfo {author} {\bibfnamefont
  {O.}~\bibnamefont {Szekely}}, \bibinfo {author} {\bibfnamefont
  {A.}~\bibnamefont {Ginsburg}}, \bibinfo {author} {\bibfnamefont
  {J.}~\bibnamefont {Mosenkis}}, \bibinfo {author} {\bibfnamefont
  {V.}~\bibnamefont {Guralnick}}, \bibinfo {author} {\bibfnamefont
  {Y.}~\bibnamefont {Dan}},  \emph {et~al.},\ }\href@noop {} {\bibfield
  {journal} {\bibinfo  {journal} {The Journal of Physical Chemistry B}\
  }\textbf {\bibinfo {volume} {115}},\ \bibinfo {pages} {14501} (\bibinfo
  {year} {2011})}\BibitemShut {NoStop}%
\bibitem [{\citenamefont {Pan}\ \emph {et~al.}(2008)\citenamefont {Pan},
  \citenamefont {Tristram-Nagle}, \citenamefont {Kucerka},\ and\ \citenamefont
  {Nagle}}]{Pan2008}%
  \BibitemOpen
  \bibfield  {author} {\bibinfo {author} {\bibfnamefont {J.}~\bibnamefont
  {Pan}}, \bibinfo {author} {\bibfnamefont {S.}~\bibnamefont {Tristram-Nagle}},
  \bibinfo {author} {\bibfnamefont {N.}~\bibnamefont {Kucerka}}, \ and\
  \bibinfo {author} {\bibfnamefont {J.~F.}\ \bibnamefont {Nagle}},\ }\href@noop
  {} {\bibfield  {journal} {\bibinfo  {journal} {Biophysical journal}\ }\textbf
  {\bibinfo {volume} {94}},\ \bibinfo {pages} {117} (\bibinfo {year}
  {2008})}\BibitemShut {NoStop}%
\bibitem [{\citenamefont {Lakshminarayanaiah}\ and\ \citenamefont
  {Murayama}(1975)}]{Lakshminarayanaiah1975}%
  \BibitemOpen
  \bibfield  {author} {\bibinfo {author} {\bibfnamefont {N.}~\bibnamefont
  {Lakshminarayanaiah}}\ and\ \bibinfo {author} {\bibfnamefont
  {K.}~\bibnamefont {Murayama}},\ }\href@noop {} {\bibfield  {journal}
  {\bibinfo  {journal} {Journal of Membrane Biology}\ }\textbf {\bibinfo
  {volume} {23}},\ \bibinfo {pages} {279} (\bibinfo {year} {1975})}\BibitemShut
  {NoStop}%
\bibitem [{\citenamefont {Kim}\ \emph {et~al.}(2014)\citenamefont {Kim},
  \citenamefont {Huang},\ and\ \citenamefont
  {Spector}}]{kim2014phosphatidylserine}%
  \BibitemOpen
  \bibfield  {author} {\bibinfo {author} {\bibfnamefont {H.-Y.}\ \bibnamefont
  {Kim}}, \bibinfo {author} {\bibfnamefont {B.~X.}\ \bibnamefont {Huang}}, \
  and\ \bibinfo {author} {\bibfnamefont {A.~A.}\ \bibnamefont {Spector}},\
  }\href@noop {} {\bibfield  {journal} {\bibinfo  {journal} {Progress in lipid
  research}\ }\textbf {\bibinfo {volume} {56}},\ \bibinfo {pages} {1} (\bibinfo
  {year} {2014})}\BibitemShut {NoStop}%
\bibitem [{\citenamefont {Johnston}\ and\ \citenamefont
  {Wu}(1994)}]{johnston1994foundations}%
  \BibitemOpen
  \bibfield  {author} {\bibinfo {author} {\bibfnamefont {D.}~\bibnamefont
  {Johnston}}\ and\ \bibinfo {author} {\bibfnamefont {S.~M.-S.}\ \bibnamefont
  {Wu}},\ }\href@noop {} {\emph {\bibinfo {title} {Foundations of cellular
  neurophysiology}}}\ (\bibinfo  {publisher} {MIT press},\ \bibinfo {year}
  {1994})\BibitemShut {NoStop}%
\bibitem [{\citenamefont {Dudel}\ \emph {et~al.}(2012)\citenamefont {Dudel},
  \citenamefont {Janig},\ and\ \citenamefont
  {Zimmermann}}]{dudel2012fundamentals}%
  \BibitemOpen
  \bibfield  {author} {\bibinfo {author} {\bibfnamefont {J.}~\bibnamefont
  {Dudel}}, \bibinfo {author} {\bibfnamefont {W.}~\bibnamefont {Janig}}, \ and\
  \bibinfo {author} {\bibfnamefont {M.}~\bibnamefont {Zimmermann}},\
  }\href@noop {} {\emph {\bibinfo {title} {Fundamentals of neurophysiology}}}\
  (\bibinfo  {publisher} {Springer Science \& Business Media},\ \bibinfo {year}
  {2012})\BibitemShut {NoStop}%
\bibitem [{\citenamefont {Angle}\ \emph {et~al.}(2015)\citenamefont {Angle},
  \citenamefont {Cui},\ and\ \citenamefont {Melosh}}]{Angle2015}%
  \BibitemOpen
  \bibfield  {author} {\bibinfo {author} {\bibfnamefont {M.~R.}\ \bibnamefont
  {Angle}}, \bibinfo {author} {\bibfnamefont {B.}~\bibnamefont {Cui}}, \ and\
  \bibinfo {author} {\bibfnamefont {N.~A.}\ \bibnamefont {Melosh}},\ }\href
  {\doibase https://doi.org/10.1016/j.conb.2015.03.014} {\bibfield  {journal}
  {\bibinfo  {journal} {Current Opinion in Neurobiology}\ }\textbf {\bibinfo
  {volume} {32}},\ \bibinfo {pages} {132 } (\bibinfo {year}
  {2015})}\BibitemShut {NoStop}%
\bibitem [{\citenamefont {Duan}\ \emph {et~al.}(2013)\citenamefont {Duan},
  \citenamefont {Fu}, \citenamefont {Liu},\ and\ \citenamefont
  {Lieber}}]{Duan2013}%
  \BibitemOpen
  \bibfield  {author} {\bibinfo {author} {\bibfnamefont {X.}~\bibnamefont
  {Duan}}, \bibinfo {author} {\bibfnamefont {T.-M.}\ \bibnamefont {Fu}},
  \bibinfo {author} {\bibfnamefont {J.}~\bibnamefont {Liu}}, \ and\ \bibinfo
  {author} {\bibfnamefont {C.~M.}\ \bibnamefont {Lieber}},\ }\href {\doibase
  https://doi.org/10.1016/j.nantod.2013.05.001} {\bibfield  {journal} {\bibinfo
   {journal} {Nano Today}\ }\textbf {\bibinfo {volume} {8}},\ \bibinfo {pages}
  {351 } (\bibinfo {year} {2013})}\BibitemShut {NoStop}%
\bibitem [{\citenamefont {Ferguson}\ \emph {et~al.}(2012)\citenamefont
  {Ferguson}, \citenamefont {Boldt}, \citenamefont {Puhl}, \citenamefont
  {Stigen}, \citenamefont {Jackson}, \citenamefont {Crisp}, \citenamefont
  {Mesce}, \citenamefont {Netoff},\ and\ \citenamefont
  {Redish}}]{Ferguson2012}%
  \BibitemOpen
  \bibfield  {author} {\bibinfo {author} {\bibfnamefont {J.~E.}\ \bibnamefont
  {Ferguson}}, \bibinfo {author} {\bibfnamefont {C.}~\bibnamefont {Boldt}},
  \bibinfo {author} {\bibfnamefont {J.~G.}\ \bibnamefont {Puhl}}, \bibinfo
  {author} {\bibfnamefont {T.~W.}\ \bibnamefont {Stigen}}, \bibinfo {author}
  {\bibfnamefont {J.~C.}\ \bibnamefont {Jackson}}, \bibinfo {author}
  {\bibfnamefont {K.~M.}\ \bibnamefont {Crisp}}, \bibinfo {author}
  {\bibfnamefont {K.~A.}\ \bibnamefont {Mesce}}, \bibinfo {author}
  {\bibfnamefont {T.~I.}\ \bibnamefont {Netoff}}, \ and\ \bibinfo {author}
  {\bibfnamefont {A.~D.}\ \bibnamefont {Redish}},\ }\href@noop {} {\bibfield
  {journal} {\bibinfo  {journal} {Nanomedicine}\ }\textbf {\bibinfo {volume}
  {7}},\ \bibinfo {pages} {847} (\bibinfo {year} {2012})}\BibitemShut {NoStop}%
\bibitem [{\citenamefont {Lacoste}\ \emph {et~al.}(2009)\citenamefont
  {Lacoste}, \citenamefont {Menon}, \citenamefont {Bazant},\ and\ \citenamefont
  {Joanny}}]{Lacoste2009}%
  \BibitemOpen
  \bibfield  {author} {\bibinfo {author} {\bibfnamefont {D.}~\bibnamefont
  {Lacoste}}, \bibinfo {author} {\bibfnamefont {G.}~\bibnamefont {Menon}},
  \bibinfo {author} {\bibfnamefont {M.}~\bibnamefont {Bazant}}, \ and\ \bibinfo
  {author} {\bibfnamefont {J.}~\bibnamefont {Joanny}},\ }\href@noop {}
  {\bibfield  {journal} {\bibinfo  {journal} {The European Physical Journal E}\
  }\textbf {\bibinfo {volume} {28}},\ \bibinfo {pages} {243} (\bibinfo {year}
  {2009})}\BibitemShut {NoStop}%
\bibitem [{\citenamefont {Lacoste}\ \emph {et~al.}(2007)\citenamefont
  {Lacoste}, \citenamefont {Lagomarsino},\ and\ \citenamefont
  {Joanny}}]{Lacoste2007}%
  \BibitemOpen
  \bibfield  {author} {\bibinfo {author} {\bibfnamefont {D.}~\bibnamefont
  {Lacoste}}, \bibinfo {author} {\bibfnamefont {M.~C.}\ \bibnamefont
  {Lagomarsino}}, \ and\ \bibinfo {author} {\bibfnamefont {J.}~\bibnamefont
  {Joanny}},\ }\href@noop {} {\bibfield  {journal} {\bibinfo  {journal} {EPL
  (Europhysics Letters)}\ }\textbf {\bibinfo {volume} {77}},\ \bibinfo {pages}
  {18006} (\bibinfo {year} {2007})}\BibitemShut {NoStop}%
\bibitem [{\citenamefont {Janssen}\ and\ \citenamefont {van
  Roij}(2017)}]{janssen2017reversible}%
  \BibitemOpen
  \bibfield  {author} {\bibinfo {author} {\bibfnamefont {M.}~\bibnamefont
  {Janssen}}\ and\ \bibinfo {author} {\bibfnamefont {R.}~\bibnamefont {van
  Roij}},\ }\href@noop {} {\bibfield  {journal} {\bibinfo  {journal} {Physical
  review letters}\ }\textbf {\bibinfo {volume} {118}},\ \bibinfo {pages}
  {096001} (\bibinfo {year} {2017})}\BibitemShut {NoStop}%
\bibitem [{\citenamefont {Biesheuvel}\ \emph {et~al.}(2015)\citenamefont
  {Biesheuvel}, \citenamefont {Hamelers},\ and\ \citenamefont
  {Suss}}]{Biesheuvel2015}%
  \BibitemOpen
  \bibfield  {author} {\bibinfo {author} {\bibfnamefont {P.}~\bibnamefont
  {Biesheuvel}}, \bibinfo {author} {\bibfnamefont {H.}~\bibnamefont
  {Hamelers}}, \ and\ \bibinfo {author} {\bibfnamefont {M.}~\bibnamefont
  {Suss}},\ }\href@noop {} {\bibfield  {journal} {\bibinfo  {journal} {Colloids
  and Interface Science Communications}\ }\textbf {\bibinfo {volume} {9}},\
  \bibinfo {pages} {1} (\bibinfo {year} {2015})}\BibitemShut {NoStop}%
\bibitem [{\citenamefont {He}\ \emph {et~al.}(2018)\citenamefont {He},
  \citenamefont {Biesheuvel}, \citenamefont {Bazant},\ and\ \citenamefont
  {Hatton}}]{he2018theory}%
  \BibitemOpen
  \bibfield  {author} {\bibinfo {author} {\bibfnamefont {F.}~\bibnamefont
  {He}}, \bibinfo {author} {\bibfnamefont {P.}~\bibnamefont {Biesheuvel}},
  \bibinfo {author} {\bibfnamefont {M.~Z.}\ \bibnamefont {Bazant}}, \ and\
  \bibinfo {author} {\bibfnamefont {T.~A.}\ \bibnamefont {Hatton}},\
  }\href@noop {} {\bibfield  {journal} {\bibinfo  {journal} {Water research}\
  }\textbf {\bibinfo {volume} {132}},\ \bibinfo {pages} {282} (\bibinfo {year}
  {2018})}\BibitemShut {NoStop}%
\bibitem [{\citenamefont {Hemmatifar}\ \emph {et~al.}(2018)\citenamefont
  {Hemmatifar}, \citenamefont {Ramachandran}, \citenamefont {Liu},
  \citenamefont {Oyarzun}, \citenamefont {Bazant},\ and\ \citenamefont
  {Santiago}}]{hemmatifar2018thermodynamics}%
  \BibitemOpen
  \bibfield  {author} {\bibinfo {author} {\bibfnamefont {A.}~\bibnamefont
  {Hemmatifar}}, \bibinfo {author} {\bibfnamefont {A.}~\bibnamefont
  {Ramachandran}}, \bibinfo {author} {\bibfnamefont {K.}~\bibnamefont {Liu}},
  \bibinfo {author} {\bibfnamefont {D.~I.}\ \bibnamefont {Oyarzun}}, \bibinfo
  {author} {\bibfnamefont {M.~Z.}\ \bibnamefont {Bazant}}, \ and\ \bibinfo
  {author} {\bibfnamefont {J.~G.}\ \bibnamefont {Santiago}},\ }\href@noop {}
  {\bibfield  {journal} {\bibinfo  {journal} {Environmental science \&
  technology}\ }\textbf {\bibinfo {volume} {52}},\ \bibinfo {pages} {10196}
  (\bibinfo {year} {2018})}\BibitemShut {NoStop}%
\bibitem [{\citenamefont {Bazant}\ \emph {et~al.}(2004)\citenamefont {Bazant},
  \citenamefont {Thornton},\ and\ \citenamefont {Ajdari}}]{Bazant2004}%
  \BibitemOpen
  \bibfield  {author} {\bibinfo {author} {\bibfnamefont {M.~Z.}\ \bibnamefont
  {Bazant}}, \bibinfo {author} {\bibfnamefont {K.}~\bibnamefont {Thornton}}, \
  and\ \bibinfo {author} {\bibfnamefont {A.}~\bibnamefont {Ajdari}},\ }\href
  {\doibase 10.1103/PhysRevE.70.021506} {\bibfield  {journal} {\bibinfo
  {journal} {Physical review E}\ }\textbf {\bibinfo {volume} {70}},\ \bibinfo
  {pages} {021056} (\bibinfo {year} {2004})}\BibitemShut {NoStop}%
\bibitem [{\citenamefont {{Janssen}}(2019)}]{Janssen2019}%
  \BibitemOpen
  \bibfield  {author} {\bibinfo {author} {\bibfnamefont {M.}~\bibnamefont
  {{Janssen}}},\ }\href@noop {} {\bibfield  {journal} {\bibinfo  {journal}
  {arXiv e-prints}\ ,\ \bibinfo {eid} {arXiv:1907.06894}} (\bibinfo {year}
  {2019})},\ \Eprint {http://arxiv.org/abs/1907.06894} {arXiv:1907.06894
  [physics.chem-ph]} \BibitemShut {NoStop}%
\bibitem [{\citenamefont {Ziebert}\ and\ \citenamefont
  {Lacoste}(2011)}]{ziebert2011}%
  \BibitemOpen
  \bibfield  {author} {\bibinfo {author} {\bibfnamefont {F.}~\bibnamefont
  {Ziebert}}\ and\ \bibinfo {author} {\bibfnamefont {D.}~\bibnamefont
  {Lacoste}},\ }in\ \href@noop {} {\emph {\bibinfo {booktitle} {Advances in
  planar lipid bilayers and liposomes}}},\ Vol.~\bibinfo {volume} {14}\
  (\bibinfo  {publisher} {Elsevier},\ \bibinfo {year} {2011})\ pp.\ \bibinfo
  {pages} {63--95}\BibitemShut {NoStop}%
\bibitem [{\citenamefont {Weast}\ \emph {et~al.}(1988)\citenamefont {Weast},
  \citenamefont {Astle}, \citenamefont {Beyer} \emph {et~al.}}]{crc}%
  \BibitemOpen
  \bibfield  {author} {\bibinfo {author} {\bibfnamefont {R.~C.}\ \bibnamefont
  {Weast}}, \bibinfo {author} {\bibfnamefont {M.~J.}\ \bibnamefont {Astle}},
  \bibinfo {author} {\bibfnamefont {W.~H.}\ \bibnamefont {Beyer}},  \emph
  {et~al.},\ }\href@noop {} {\emph {\bibinfo {title} {CRC handbook of chemistry
  and physics}}},\ Vol.~\bibinfo {volume} {69}\ (\bibinfo  {publisher} {CRC
  press Boca Raton, FL},\ \bibinfo {year} {1988})\BibitemShut {NoStop}%
\bibitem [{\citenamefont {Jacobs}(2013)}]{Jacobs2013}%
  \BibitemOpen
  \bibfield  {author} {\bibinfo {author} {\bibfnamefont {P.}~\bibnamefont
  {Jacobs}},\ }\href {https://books.google.be/books?id=hvA7DQAAQBAJ} {\emph
  {\bibinfo {title} {Thermodynamics}}}\ (\bibinfo  {publisher} {World
  Scientific Publishing Company},\ \bibinfo {year} {2013})\BibitemShut
  {NoStop}%
\bibitem [{\citenamefont {Carvalho-de Souza}\ \emph {et~al.}(2015)\citenamefont
  {Carvalho-de Souza}, \citenamefont {Treger}, \citenamefont {Dang},
  \citenamefont {Kent}, \citenamefont {Pepperberg},\ and\ \citenamefont
  {Bezanilla}}]{carvalho2015}%
  \BibitemOpen
  \bibfield  {author} {\bibinfo {author} {\bibfnamefont {J.~L.}\ \bibnamefont
  {Carvalho-de Souza}}, \bibinfo {author} {\bibfnamefont {J.~S.}\ \bibnamefont
  {Treger}}, \bibinfo {author} {\bibfnamefont {B.}~\bibnamefont {Dang}},
  \bibinfo {author} {\bibfnamefont {S.~B.}\ \bibnamefont {Kent}}, \bibinfo
  {author} {\bibfnamefont {D.~R.}\ \bibnamefont {Pepperberg}}, \ and\ \bibinfo
  {author} {\bibfnamefont {F.}~\bibnamefont {Bezanilla}},\ }\href@noop {}
  {\bibfield  {journal} {\bibinfo  {journal} {Neuron}\ }\textbf {\bibinfo
  {volume} {86}},\ \bibinfo {pages} {207} (\bibinfo {year} {2015})}\BibitemShut
  {NoStop}%
\bibitem [{\citenamefont {Martino}\ \emph {et~al.}(2015)\citenamefont
  {Martino}, \citenamefont {Feyen}, \citenamefont {Porro}, \citenamefont
  {Bossio}, \citenamefont {Zucchetti}, \citenamefont {Ghezzi}, \citenamefont
  {Benfenati}, \citenamefont {Lanzani},\ and\ \citenamefont
  {Antognazza}}]{martino2015}%
  \BibitemOpen
  \bibfield  {author} {\bibinfo {author} {\bibfnamefont {N.}~\bibnamefont
  {Martino}}, \bibinfo {author} {\bibfnamefont {P.}~\bibnamefont {Feyen}},
  \bibinfo {author} {\bibfnamefont {M.}~\bibnamefont {Porro}}, \bibinfo
  {author} {\bibfnamefont {C.}~\bibnamefont {Bossio}}, \bibinfo {author}
  {\bibfnamefont {E.}~\bibnamefont {Zucchetti}}, \bibinfo {author}
  {\bibfnamefont {D.}~\bibnamefont {Ghezzi}}, \bibinfo {author} {\bibfnamefont
  {F.}~\bibnamefont {Benfenati}}, \bibinfo {author} {\bibfnamefont
  {G.}~\bibnamefont {Lanzani}}, \ and\ \bibinfo {author} {\bibfnamefont
  {M.~R.}\ \bibnamefont {Antognazza}},\ }\href@noop {} {\bibfield  {journal}
  {\bibinfo  {journal} {Scientific reports}\ }\textbf {\bibinfo {volume} {5}},\
  \bibinfo {pages} {8911} (\bibinfo {year} {2015})}\BibitemShut {NoStop}%
\bibitem [{\citenamefont {Hotka}\ and\ \citenamefont
  {Zahradnik}(2014)}]{hotka2014}%
  \BibitemOpen
  \bibfield  {author} {\bibinfo {author} {\bibfnamefont {M.}~\bibnamefont
  {Hotka}}\ and\ \bibinfo {author} {\bibfnamefont {I.}~\bibnamefont
  {Zahradnik}},\ }\href@noop {} {\bibfield  {journal} {\bibinfo  {journal}
  {Biophysical Journal}\ }\textbf {\bibinfo {volume} {106}},\ \bibinfo {pages}
  {121a} (\bibinfo {year} {2014})}\BibitemShut {NoStop}%
\bibitem [{\citenamefont {Shapiro}\ \emph {et~al.}(2012)\citenamefont
  {Shapiro}, \citenamefont {Homma}, \citenamefont {Villarreal}, \citenamefont
  {Richter},\ and\ \citenamefont {Bezanilla}}]{Shapiro2012}%
  \BibitemOpen
  \bibfield  {author} {\bibinfo {author} {\bibfnamefont {M.~G.}\ \bibnamefont
  {Shapiro}}, \bibinfo {author} {\bibfnamefont {K.}~\bibnamefont {Homma}},
  \bibinfo {author} {\bibfnamefont {S.}~\bibnamefont {Villarreal}}, \bibinfo
  {author} {\bibfnamefont {C.~P.}\ \bibnamefont {Richter}}, \ and\ \bibinfo
  {author} {\bibfnamefont {F.}~\bibnamefont {Bezanilla}},\ }\href {\doibase
  10.1038/ncomms1742} {\bibfield  {journal} {\bibinfo  {journal} {Nature
  Communications}\ }\textbf {\bibinfo {volume} {3}},\ \bibinfo {pages} {310}
  (\bibinfo {year} {2012})}\BibitemShut {NoStop}%
\end{thebibliography}%
\section{Appendix}
\subsection{Fast equilibration of double layers}\label{sec:fast-eq}
In this work, we describe ions and the electric potential in the diffuse layers at equilibrium, with the Boltzmann distribution (Eqs. \eqref{eq:PBi} and \eqref{eq:PBo}). This equilibrium assumption decouples the problem from time dependence for relaxation of the double layers and simplifies it to a single dependence on the membrane potential $V_{\mathrm{m}}$. However, in reality, $V_{\mathrm{m}}$ is time-dependent. We must thus verify that the double layers reach their equilibrium conformation (ion concentration versus position) significantly faster than the variation of $V_{\mathrm{m}}$ during the course of the action potential. Two time scales can characterize the relaxation of diffuse layers: diffusion $\tau_D={L^2}/{D}$ and charge density relaxation $\tau_L={\lambda_D^2}/{D}$, where L is the characteristic length, D the characteristic diffusivity, and $\lambda_D$ the Debye length. At low voltages, \citet{Bazant2004} have shown that the primary time scale for diffuse charges dynamics in a non-Faradaic electrochemical cell is the harmonic mean between these time constants, that is: $\tau_C=\lambda_D\:L/D$. Interestingly, a recent study by \citet{Janssen2019} has shown that the easily obtainable $\tau_C$ decently describes double layer relaxation also in cylindrical geometry, the relevant geometry for neurons.
By analogy with the electrochemical cell, the membrane can be seen as sandwiched between two electrodes separated by a distance $L$, and on which a difference of potential is applied \cite{Lacoste2009,ziebert2011}.\\
In physiological conditions (ionic strength = 150 mM) and at 0$\degree$C, the Debye length is less than 1 nm:
\begin{equation}
\lambda_D = \sqrt{\frac{RT\:\varepsilon}{F^2 \sum{z_j^2\:c_j}}} = \: 0.6 \: \mathrm{nm}.
\end{equation}
In the garfish olfactory nerve, one of the smallest axons studied, the characteristic length is given by the radius of the cell, $L/2 = 0.25$ $\SI{}{\micro\meter}$. Taking $D= 10^{-9}$ m$^{2}$ $s^{-1}$ we find $\tau_C = 0.3$ $\SI{}{\micro\second}$. In the giant squid axon, one of the largest axons ever studied, we have $L/2 = 200$ $\SI{}{\micro\meter}$, and thus $\tau_C = 240$ $\SI{}{\micro\second}$.  
These characteristic times for relaxation are small compared to the time scale of compound action potentials, which is 10 to 100 ms \cite{Howarth1979}. While in the nonlinear regime (where the double layer potential across each side is much larger than the thermal voltage $\sim$24 mV) the time scale for bulk salt diffusion $\tau_D=L^2/D$ can be important, the leading order dynamics occur at the $\tau_C$ scale. Therefore, we neglect any double layer relaxation effect and treat them at equilibrium.  \\
\subsection{Fast equilibration of temperature inside the nerve}\label{sec:fast-eq-T}
Another important assumption we make is the fast heat equilibration between the cell membrane and the rest of the nerve. The validity of this assumption can be verified based on scaling analysis on the heat equation, which yields a characteristic time for heat diffusion:
\begin{equation}
    \tau_{\mathrm{h}} = \frac{\rho\:c_{\mathrm{p}}\:L^2}{K} 
\end{equation} where $\rho$ is the mass density, $c_{\mathrm{p}}$ the specific heat capacity, $L$ the characteristic length and $K$ the thermal conductivity. With $L = 0.125 \:\SI{}{\micro\meter}$, $\rho = 5.8$ g cm$^{-3}$, $c_{\mathrm{p}} = 3.6$ J (g.K)$^{-1}$ \cite{Howarth1979}, and using the thermal conductivity of water at 0$\degree$C, $K=0.0056$ J (K cm s)$^{-1}$ \cite{crc}, we find $\tau_{\mathrm{h}} \sim 0.6$ \micro{s}.
Thus, the heat equilibration in the interior of the small nerve fibers in which the heat experiments were conducted \cite{Howarth1979} is on the order of microseconds, which is much quicker than the timescale of the action potential in these nerves (10 to 100 ms) \cite{Howarth1979}. 
The nerve is solely composed of densely-packed fibers, such that the mean distance between fibers is smaller than the size of a single fiber \cite{easton1971}. Thus, we assume that the whole nerve volume is at thermal equilibrium during the action potential, and calculate the rise of the nerve's temperature as proportional to the amount of heat dissipated:
\begin{equation}
    \Delta{T} = -\frac{\Delta{Q}}{c_{\mathrm{p}}}\:A_{\mathrm{m}},
\end{equation}
where $A_{\mathrm{m}} = 65000$ cm$^2$ g$^{-1}$ is the total membrane area per mass of nerve \cite{Howarth1979}.
\subsection{The Condenser Theory from Maxwell's equation for Ampere's law}\label{sec:Maxwell}
Heat production and field energy can be unified by Maxwell's equation for Ampere's law \cite{Jacobs2013},
\begin{equation}
E\:(\nabla\times{H}) = E\:{J} + E\:{\frac{\partial{D}}{\partial{t}}},
\end{equation}
where $J$ is the current density (A m$^{-2}$), $H$ is the magnetic field and $D$ the displacement.  -$E\:{J}$ is the energy dissipated by an ionic current $J$ in an electric field $E$ as heat per unit of volume and time \cite{Jacobs2013}, noted $\dot{Q}_{\mathrm{V}}$ (negative when heat is dissipated).
Neglecting magnetic contributions, and in one dimension, we find 
\begin{equation}\label{eq:maxwell}
\dot{Q}_{\mathrm{V}} =- E\:J = E\:{\frac{\partial{D}}{\partial{t}}}.
\end{equation}
This relationship offers another way to understand the Condenser Theory: during the action potential, positive charges (\ce{Na+} and \ce{K+} ions) move through ion channels of the membrane \cite{Hille2001}, first in the same direction as the electric field ($- E.J < 0$, in one dimension), then against the electric field ($- E.J > 0$), which results in heat production ($\dot{Q}_{\mathrm{V}} < 0$) and then heat absorption ($\dot{Q}_{\mathrm{V}} > 0$).
Let's consider the situation where the field is 0 at time zero and $E$ at time $t$. If the dielectric medium is linear ($D = \varepsilon.E$), the heat absorbed between time $0$ and $t$ is equal to the field energy at time $t$:
\begin{equation}
\begin{split}
Q_{\mathrm{V}} &= \int_0^t{E\:{\frac{\partial{D}}{\partial{t}}}}\:dt = \frac{1}{2}\int_0^t{\Big(E\:{\frac{\partial{D}}{\partial{t}}}+\frac{\partial{E}}{\partial{t}}\:{D}\Big)}\:dt \\ &= \frac{1}{2} E\:{D}.
\end{split}
\end{equation}
The second equality holds for any linear dielectric medium, whereas the third is given by the fundamental theorem of calculus. Finally, integrating $Q_{\mathrm{V}}$ over the membrane domain $\Omega_{\mathrm{M}}$ gives the heat dissipated per unit of membrane surface area (J/m$^2$):
\begin{equation}\label{eq:free_el_energy2}
  Q = \int^{+\infty}_{-\infty}{\int_0^t{E\:{\frac{\partial{D}}{\partial{t}}}}\:dt \:dx = \int^{+\infty}_{-\infty} \frac{1}{2} E\:D}\:dx.
\end{equation}
This last expression gives the amount of heat that is dissipated upon release of the electric free energy of a dielectric medium \cite{Overbeek1990,Jackson1999}. 

\subsection{Derivation of the electric free energy of the membrane capacitance and of double layers}\label{sec:free_energy_derivation}
In this section we split the free energy of the $\Omega_{\mathrm{M}}$ domain (given by Eq. \eqref{eq:free_el_energy} or \eqref{eq:free_el_energy2}) into a contribution from the membrane capacitance and one form the double layers. This separation is necessary to entropy calculations (Section \ref{sec:entropy}). We start by expanding the field energy expression over the internal diffuse layer, the membrane, and the external diffuse layer:
\begin{equation}\label{eq:F2}
\begin{split}
F_{\mathrm{el}} =& \frac{1}{2}\bigg[ \varepsilon \int_{-\infty}^{-\delta}{\bigg(\frac{d{\phi_{\mathrm{i}}(x)}}{d{x}}}\bigg)^2dx+ \varepsilon_{\mathrm{m}} \int_{-\delta}^{0}{\bigg(\frac{d{\phi_{\mathrm{m}}(x)}}{d{x}}}\bigg)^2 dx \\
&+ \varepsilon \int_{0}^{+\infty}{\bigg(\frac{d{\phi_{\mathrm{o}}(x)}}{d{x}}}\bigg)^2dx \bigg].\\
\end{split}
\end{equation}
Assuming the density of free charges to be zero at any point $]-\delta;0[$ inside the membrane, Poisson's equation requires the field to be constant across the membrane, such that $\frac{d{\phi_{\mathrm{m}}(x)}}{d{x}}$ = $\frac{d{\phi_{\mathrm{m}}(0)}}{d{x}}$. The second term in the RHS of Eq. \eqref{eq:F2}, then simplifies to
\begin{equation}
\begin{split}
\varepsilon_{\mathrm{m}} \int_{-\delta}^{0}\bigg(\frac{d{\phi_{\mathrm{m}}(x)}}{d{x}}\bigg)^2 dx &= \varepsilon_{\mathrm{m}} \int_{-\delta}^{0} \bigg(\frac{d\phi_{\mathrm{m}}(0)}{dx}\bigg)^2 dx \\
&= \varepsilon_{\mathrm{m}} \bigg(\frac{q}{\varepsilon_{\mathrm{m}}}\bigg)^2  \int_{-\delta}^{0}{}dx \\
&= \frac{q^2}{c_{\mathrm{m}}}.\\
\end{split}
\end{equation}
The second equality is given by boundary conditions for the electric field, Eq. \eqref{eq:24}. Next, integration by parts of the first and third terms in Eq. \eqref{eq:F2} gives
\begin{equation} \label{eq:F3}
\begin{split}
F_{\mathrm{el}} =&  \frac{1}{2}\bigg( \varepsilon\:\left[{\frac{d{\phi_{\mathrm{i}}}}{d{x}}\:\phi_{\mathrm{i}}}\right]^{-\delta}_{-\infty} - \varepsilon \int_{-\infty}^{-\delta}{\phi_{\mathrm{i}}\:\frac{d^2{\phi_{\mathrm{i}}}}{d{x^2}}} dx +\frac{q^2}{c_{\mathrm{m}}}\\ 
&+  								\varepsilon\:\left[{\frac{d{\phi_{\mathrm{o}}}}{d{x}}\:\phi_{\mathrm{o}}}\right]^{+\infty}_{0}  - \varepsilon \int^{+\infty}_{0}{\phi_{\mathrm{o}}\:\frac{d^2{\phi_{\mathrm{o}}}}{d{x^2}}}dx\bigg).
\end{split}
\end{equation}
Remembering that the potential is constant at $-\infty$ and $+\infty$, using Eqs. \eqref{eq:18} and \eqref{eq:19} and Poisson's equation, Eq. \eqref{eq:F3} can be rearranged into
\begin{equation}\label{eq:F4}
\begin{split}
F_{\mathrm{el}} =&  \frac{1}{2}\bigg(  \int_{-\infty}^{-\delta}{\rho_{\mathrm{i}}\:\phi_{\mathrm{i}}\:} dx + \Big(\sigma_{\mathrm{i}} +q\Big)\:\phi_{\mathrm{i}}(-\delta)\\ &+ \int^{+\infty}_{0}{\rho_{\mathrm{o}}\:\phi_{\mathrm{o}}\:}dx  + \Big(\sigma_{\mathrm{o}} - q\Big)\:\phi_{\mathrm{o}}(0) +\frac{q^2}{c_{\mathrm{m}}} \bigg).\\
\end{split}
\end{equation}
As $q = - c_{\mathrm{m}}(\phi_{\mathrm{i}}(-\delta) - \phi_{\mathrm{o}}(0))$ by Eq. \eqref{eq:q}, this last expression simplifies to
\begin{equation}
\begin{split}
F_{\mathrm{el}} =& \frac{1}{2}\bigg( \int_{-\infty}^{-\delta}{\rho_{\mathrm{i}}\:\phi_{\mathrm{i}}\:} dx + \sigma_{\mathrm{i}}\:\phi_{\mathrm{i}}(-\delta)\\
&+ \int^{+\infty}_{0}{\rho_{\mathrm{o}}\:\phi_{\mathrm{o}}\:}dx + \sigma_{\mathrm{o}}\:\phi_{\mathrm{o}}(0)\bigg).
\end{split}
\end{equation}
By applying the substitution $\phi_{\mathrm{i}} = (\phi_{\mathrm{i}} - V_{\mathrm{m}}) + V_{\mathrm{m}}$ and using Eqs. \eqref{eq:rho_i} and \eqref{eq:q}, we separate the free energy in the diffuse layers $ (F^{\mathrm{DL}}_{\mathrm{el}})$ from the one in the membrane $(F^{\mathrm{m}}_{\mathrm{el}})$:
\begin{equation}
\begin{split}
F^{\mathrm{DL}}_{\mathrm{el}} =& \frac{1}{2}\bigg( \int_{-\infty}^{-\delta}{\rho_{\mathrm{i}}(\phi_{\mathrm{i}} - V_{\mathrm{m}})} dx + \sigma_{\mathrm{i}} \:(\phi_{\mathrm{i}}(-\delta) - V_{\mathrm{m}}) \\&+ \int^{+\infty}_{0}{\rho_{\mathrm{o}}\:\phi_{\mathrm{o}}\:}dx + \sigma_{\mathrm{o}}\:\phi_{\mathrm{o}}(0)\bigg),
\end{split}
\end{equation}
and
\begin{equation}
F^{\mathrm{m}}_{\mathrm{el}} =  \frac{1}{2}\:c_{\mathrm{m}}\: \phi_{\mathrm{t}}\:V_{\mathrm{m}},
\end{equation}
with $F_{\mathrm{el}} = F^{\mathrm{DL}}_{\mathrm{el}} + F^{\mathrm{m}}_{\mathrm{el}}$. 
\subsection{Derivation of entropy changes in the membrane}\label{sec:entropy_derivation}
Within the Condenser Theory, it was suggested that entropy changes in the lipid bilayer accompanying the action potential could account for up to 4 times the free energy changes \cite{Howarth1979,Ritchie1973}. The ratio between the entropy-related energy and the free energy was calculated as
\begin{equation}\label{eq:entropy-free_energy2}
\frac{T\Delta{S}}{\Delta{F}} = \frac{T}{C}\:\frac{\partial{C}}{\partial{T}}
\end{equation}
where $C$ is the membrane capacitance. However, no full derivation of this equation is reported in the literature. Here we will show under which conditions Eq. \ref{eq:entropy-free_energy2} can be derived, based on the demonstration \citet{Frohlich1968}. For simplicity, we will consider a membrane without any surface charges, holding a charge $q$ on each side (here $q$ is the capacitive charge in C), across a potential difference $V$. This simplification will ease notations without modifying the relation between the membrane's entropy and electric free energy studied here. We start by considering the change of internal energy $dU$ (in J) following an increment of charge $dq$. The first law of thermodynamics reads
\begin{align*}
dU &= dQ + dW \\
&= dQ + V \:dq
\end{align*}
where $dQ$ is the influx of heat and $dW$ is the electrical work done on the membrane. Note that the electrical work did not appear explicitly in the thermodynamic system defined in Section \ref{sec:thermo}, since no electrical work is done on $\Omega_{\mathrm{M}}$ by the surroundings (the latter is however implicitly accounted for as $\Delta{F_{\mathrm{el}}}$ in Eq. \ref{eq:dUdef}). For the purpose of this demonstration, the thermodynamic system considered here consists of the membrane only ($x\: \epsilon\: ]-\delta;0[$), which enables us to express the electrical work associated with the depolarization and repolarization of the membrane explicitly. Assuming that the capacitance $C$ remains constant with $V$ but depends on the temperature $T$, we obtain
\begin{equation}
dq = d(C.V) = dC.V + C.dV = \frac{\partial{C}}{\partial{T}}\:dT.V + C.d(V).
\end{equation}
Thus, a variation of $q$ may be due to a variation of temperature or of potential, the capacitance being assumed to remain constant for the range of physiological potentials ($ca.$ - 100 mV to + 20 mV). By taking $T$ and $V^2$ as independent variables, the first law now becomes
\begin{equation}\label{eq:firstlaw}
dQ + V^2 \frac{\partial{C}}{\partial{T}}\:dT + \frac{1}{2} C\:d(V^2)  = dU =  \frac{\partial{U}}{\partial{T}}\:dT + \frac{\partial{U}}{\partial{(V^2)}}\:d(V^2).
\end{equation}
Note that $dQ$ is not a \textit{total differential}, however $dS = dQ/T$ is one, for a reversible process \cite{Jackson1999}. This property means that a unique function $S(T,V^2)$ must exist, such that
\begin{equation}\label{eq:totaldiff}
dS = \frac{\partial{S}}{\partial{T}}\:dT + \frac{\partial{S}}{\partial{(V^2)}}\:d(V^2).
\end{equation}
Thus if there is a relation $dS = f(T,V^2).dT + g(T,V^2).dV^2$, where \textit{f} and \textit{g} are two unknown functions, the condition that $dS$ is a \textit{total differential} requires that
\begin{equation}\label{eq:partial} 
\frac{\partial{f}}{\partial{(V^2)}} = \frac{\partial{g}}{\partial{T}},
\end{equation}
as both sides of Eq. \ref{eq:partial} are equal to ${\partial^2{S}}/({\partial{T}\partial{(V^2)}})$, by Eq. \ref{eq:totaldiff}. 
Next, by substituting for $dS = dQ/T$ in Eq. \eqref{eq:firstlaw}, we find
\begin{equation}\label{eq:dS}
dS = \frac{1}{T}\Big( \frac{\partial{U}}{\partial{T}} - V^2 \frac{\partial{C}}{\partial{T}}\Big) dT +  \frac{1}{T}\Big( \frac{\partial{U}}{\partial{(V^2)}} - \frac{C}{2}\Big)d(V^2).
\end{equation}
This equation is analogous to Eq. \eqref{eq:totaldiff} and therefore, Eq. \eqref{eq:partial} becomes
\begin{equation}
\frac{\partial}{\partial{(V^2)}}\Big\{\frac{1}{T}\Big(\frac{\partial{U}}{\partial{T}} - V^2 \frac{\partial{C}}{\partial{T}}\Big)\Big\} =  \frac{\partial}{\partial{T}}\Big\{\frac{1}{T}\Big(\frac{\partial{U}}{\partial{(V^2)}} - \frac{C}{2}\Big)\Big\}.
\end{equation}
By carrying out differentiations we find (take ${\partial{C}}/{\partial{T}}$ as a constant): 
\begin{equation}
\frac{\partial{U}}{\partial{(V^2)}} = \frac{C}{2}  + \frac{1}{2}\frac{\partial{C}}{\partial{T}}T
\end{equation}
Finally, upon integration, we obtain
\begin{dmath}
\Delta{U} = \frac{1}{2} \int^{V^2}_0 \Big( C + \frac{\partial{C}}{\partial{T}}T\Big)d(V^2) 
 = \frac{1}{2}CV^2 + \frac{1}{2}\frac{\partial{C}}{\partial{T}}T\:V^2.
\end{dmath}
The first term is the free energy change $\Delta{F}$ of the capacitor, whereas the second is the energy associated with a change of entropy $T\Delta{S}$. 
Thus we have
\begin{equation}
T\Delta{S} = \Delta{U} - \Delta{F} =  \frac{1}{2}\frac{\partial{C}}{\partial{T}}T\:V^2 =  \frac{T}{C}\:\frac{\partial{C}}{\partial{T}} \Delta{F},
\end{equation}
and Eq. \eqref{eq:entropy-free-energy-m} holds, under the condition that
the membrane capacitance has a linear dependence on temperature, but is constant with potential. A number of recent studies indicate that there exists such a linear dependence and that the value of ${\partial{C}}/{\partial{T}}$ is $\sim 0.3 \%/\degree $C for a wide range of cell membranes \cite{Plaksin2018,carvalho2015,martino2015,hotka2014,Shapiro2012}. 
Why does the membrane capacitance change with temperature? 
Using the classical model for the membrane capacitance, $C = \varepsilon_{\mathrm{m}}\:{A}/{d}$, where $A$ is the membrane area, and $\delta$ the membrane thickness, the temperature dependence could be attributed to a variation of the dielectric permittivity ${\partial{\varepsilon_{\mathrm{m}}}}/{\partial{T}}$, or to a variation of the membrane dimensions (${\partial{A}}/{\partial{T}}$ and ${\partial{\delta}}/{\partial{T}}$). Dimensional changes have been measured in artificial lipid bilayers \cite{Pan2008,Kucerka2011,szekely2011effect}, and in neurons, these changes correspond to a $0.11$+-$0.03\%$/K reduction in the lipid membrane thickness and a $0.22$+-$0.03\%/\degree$C increase in the area per lipid, resulting in a linear increase of the capacitance with temperature of 0.3 $\%$/K \cite{Plaksin2018}. 
Little is known on how the membrane's dielectric permittivity varies with temperature, however ${\partial{\varepsilon_{\mathrm{m}}}}/{\partial{T}}$ has been measured for unsaturated fatty acids (the molecules composing the core of the membrane), yielding values more than one order of magnitude smaller than the combined dimensional changes \cite{Plaksin2018,Lizhi2008}.
\subsection{Supplementary plots}\label{sec:suppl_plot}
\begin{figure}[h]
\includegraphics[width=1\linewidth]{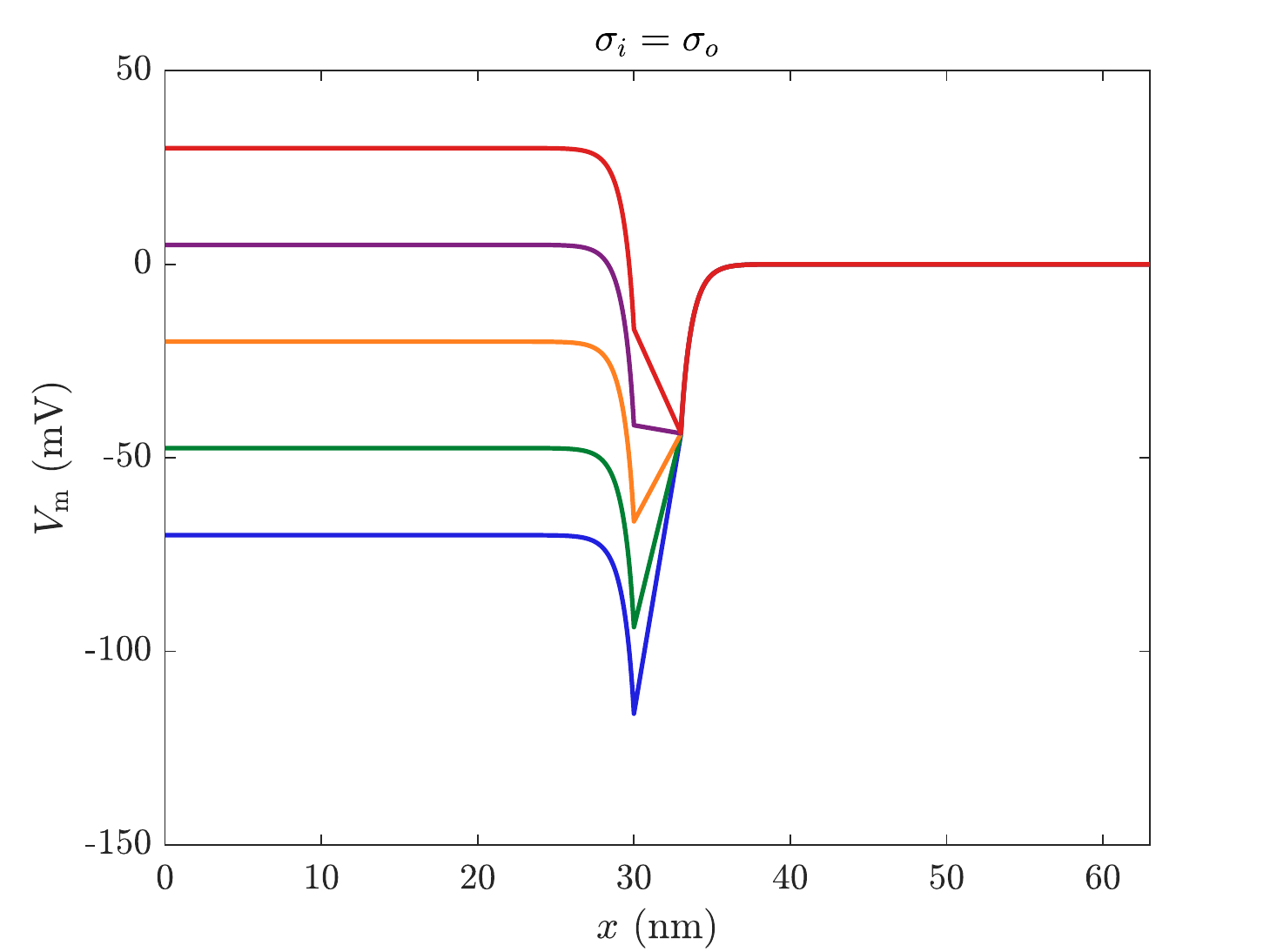}
\end{figure}
\begin{figure}[h]
\includegraphics[width=1\linewidth]{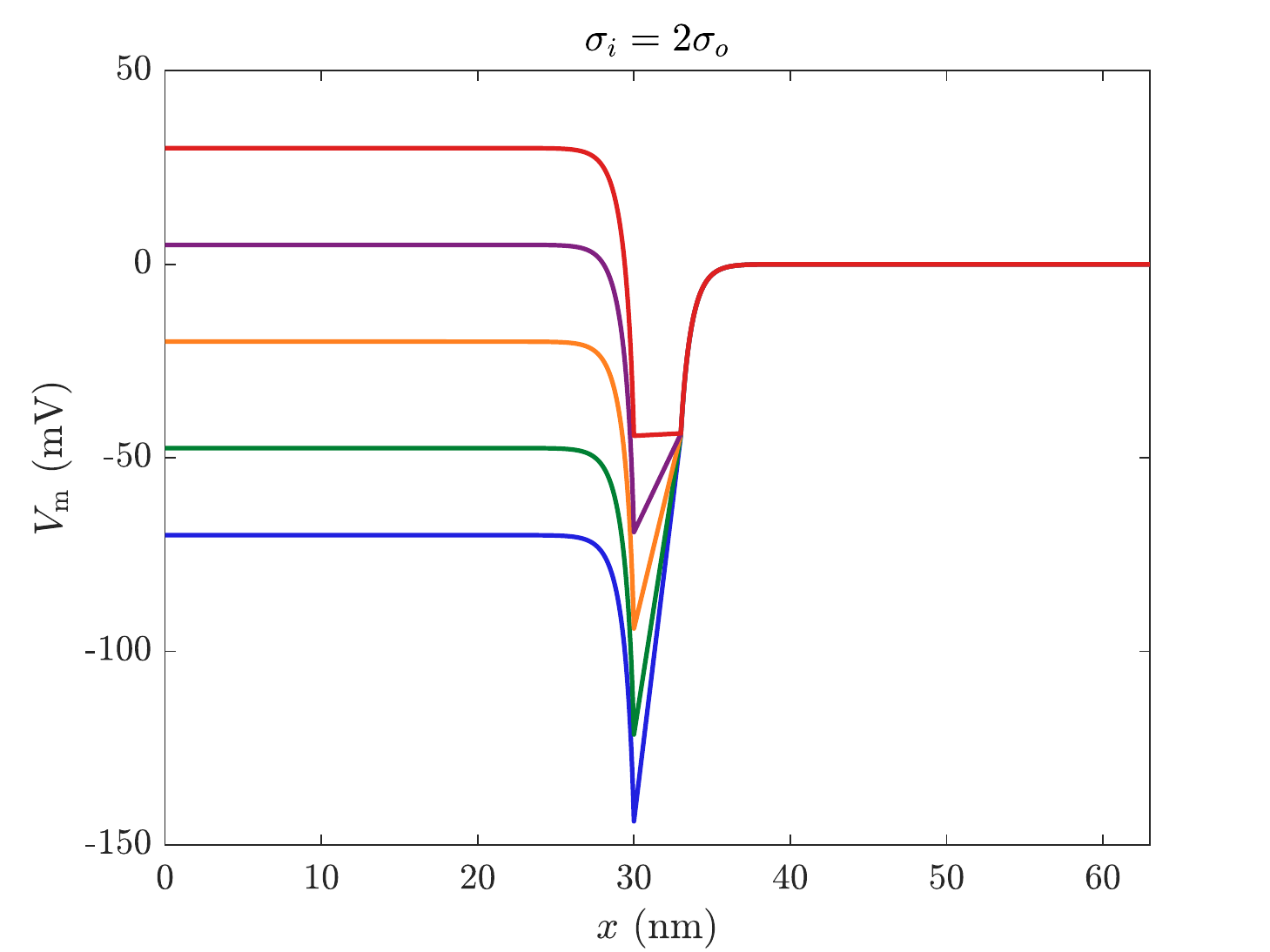}
\caption{Profiles of the electric potential across the membrane and the surrounding diffuse layers, for different values of the membrane potential $V_{\mathrm{m}}$ (-70 to +30 mV). TOP: surface charge densities are equal on each side ($\sigma_{\mathrm{i}} =  \sigma_{\mathrm{o}} = -0.05 $ C m$^{-2}$). BOTTOM: a bias of surface charges makes the internal side more negative ($\sigma_{\mathrm{i}} = $-0.1 C m$^{-2}$ and $\sigma_{\mathrm{o}} = -0.05 $ C m$^{-2}$), such that the transmembrane potential $\phi_{\mathrm{t}}$ need not to reverse when the membrane potential is positive (in purple).}
\label{fig:potential_profiles}
\end{figure}
\end{document}